\documentclass{emulateapj}

\usepackage{txfonts}
\usepackage{multirow}
\usepackage{verbatim}
\usepackage{color}
\usepackage{sidecap}
\usepackage{natbib}

\definecolor{orange}{rgb}{1,.5,0}

\def\kms{\mbox{km~s$^{-1}$}}

\newcommand{\cmd}   {~cm$^{-2}$}

\newcommand{\ndh}  {N$_2$H$^+$}
\newcommand{\ndd}  {N$_2$D$^+$}
\newcommand{\dcop} {DCO$^+$}

\slugcomment{}
\shorttitle{Starless Cores in the Pipe Nebula}
\shortauthors{Frau et al.}

\begin{document}

\title{
Young starless cores embedded in the magnetically dominated Pipe Nebula
\thanks{Based on observations carried out with the IRAM 30-m telescope. IRAM is supported by
INSU/CNRS (France), MPG (Germany), and IGN (Spain).}
}

\author{P.\ Frau\altaffilmark{1}, J.\ M.\ Girart\altaffilmark{1}, 
M.\ T.\ Beltr\'an \altaffilmark{2}, O.\ Morata \altaffilmark{3,4}, 
J.\ M.\ Masqu\'e \altaffilmark{5}, 
\\
G.\ Busquet \altaffilmark{5}, 
F.\ O.\ Alves \altaffilmark{1}, \'A.\ S\'anchez-Monge \altaffilmark{5}, 
R.\ Estalella \altaffilmark{5} and G.\ A.\ P. Franco \altaffilmark{6}}

\affil{$^1$ Institut de Ci\`encies de l'Espai (CSIC-IEEC), Campus UAB, Facultat de Ci\`encies, Torre C-5p, 08193 Bellaterra, Catalunya, Spain}
\affil{$^2$ INAF-Osservatorio Astrofisico di Arcetri, Largo Enrico Fermi 5, 50125 Firenze,Italy}
\affil{$^3$ Institute of Astronomy and Astrophysics, Academia Sinica, P.O.\ Box 23-141, Taipei 106, Taiwan}
\affil{$^4$ Department of Earth Sciences, National Taiwan Normal University, 88, Section 4, Ting-Chou Road, Taipei 11677, Taiwan.}
\affil{$^5$ Departament d'Astronomia i Meteorologia and Institut de Ci\`encies del Cosmos (IEEC-UB),
 Universitat de Barcelona, Mart{\'\i} i Franqu\`es 1,
08028 Barcelona, Catalunya, Spain}
\affil{$^6$ Departamento de F\'isica - ICEx - UFMG, Caixa Postal 702, 30.123-970, Belo Horizonte, Brazil}

\begin{abstract}

The Pipe Nebula is a massive, nearby dark molecular cloud with a low star-formation  efficiency  which makes it a  good
laboratory to study the very early stages of the  star formation process.  The Pipe Nebula is largely filamentary, and
appears to be threaded by a uniform  magnetic field at scales of few parsecs, perpendicular to its main axis. The field
is only locally perturbed in a few regions, such as the only active cluster forming core B59. The aim of this study is to
investigate primordial conditions in low-mass pre-stellar cores and how they relate to the local magnetic field in the
cloud. We used the IRAM 30-m telescope to carry out a continuum and molecular survey at 3 and 1~mm of early- and 
late-time molecules toward four selected starless cores inside the Pipe Nebula. We found that the dust continuum emission
maps trace better the densest regions than previous 2MASS extinction maps, while  2MASS extinction maps trace better the
diffuse gas. The properties of the cores derived from dust emission show average radii of $\sim$0.09~pc,  densities of
$\sim$1.3$\times10^{5}$~cm$^{-3}$, and core masses of $\sim$2.5~$M_{\odot}$.   Our results confirm that the Pipe Nebula
starless cores studied are in a very early evolutionary stage, and present a very young chemistry with different
properties that allow us to propose an evolutionary sequence. All of the cores present early-time molecular emission,
with CS detections toward all the sample. Two of them,  Cores~40 and 109, present strong late-time molecular emission.
There seems to be a correlation between the chemical evolutionary stage of the cores and the local magnetic properties
that  suggests that the evolution of the cores is ruled by a local competition  between the magnetic energy and other
mechanisms, such as turbulence.

\end{abstract}

\keywords{
ISM: individual objects: Pipe Nebula -- 
ISM: lines and bands --
ISM -- stars: formation}

\section{Introduction}\label{intro}

   \begin{figure*}
   \includegraphics[width=14cm,angle=-90]{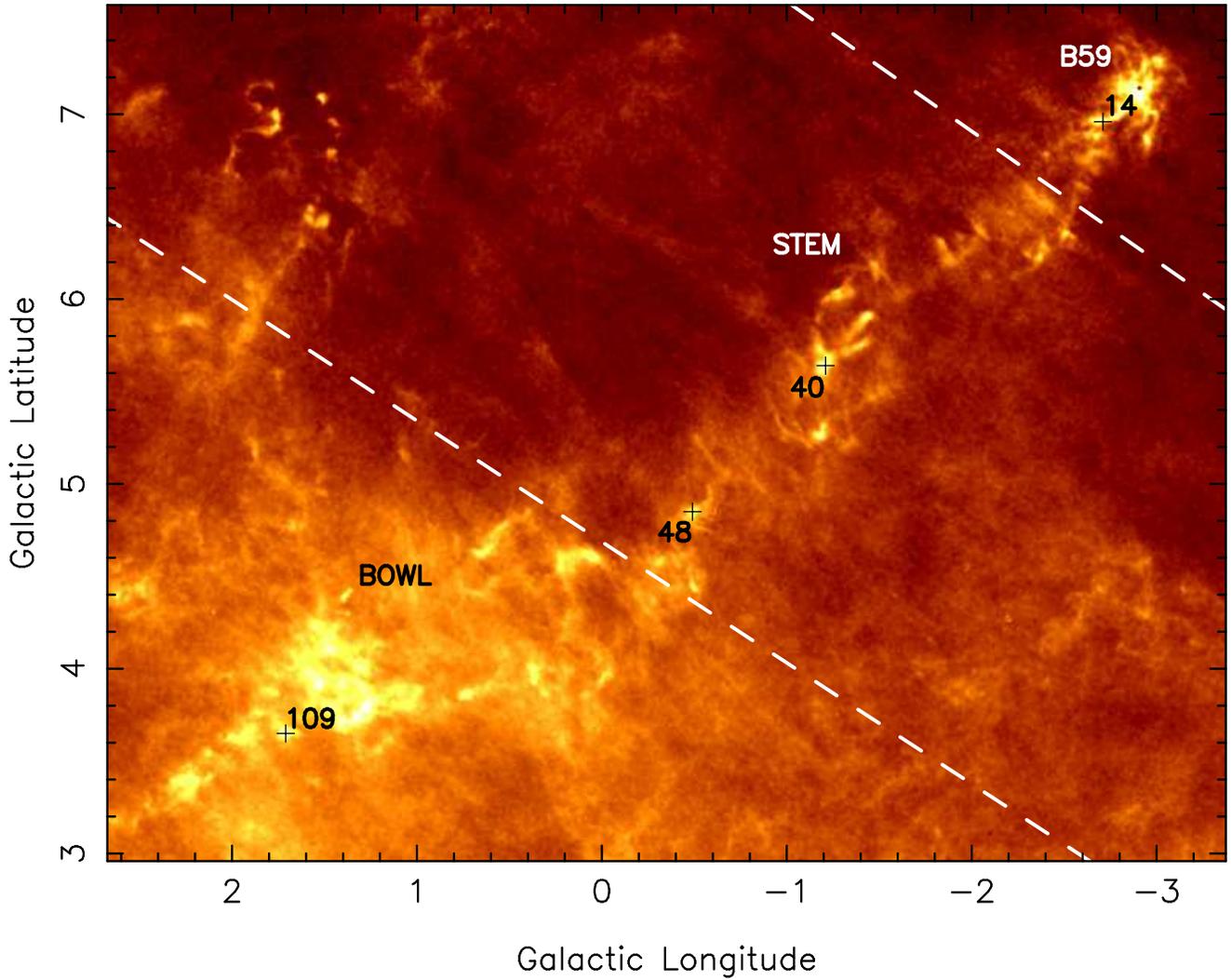}
      \caption{
Position of the observed cores plotted over the 2MASS extinction map of the Pipe
Nebula (Lombardi et al, 2006). 
The dashed lines separate the three different magnetically defined regions \citep{alves08}.
The lowest visual extinction ($A_\texttt{v}$) corresponds 
to 0.5 magnitudes. The
highest $A_\texttt{v}$ is observed toward the {\it bowl} of the Pipe and the B59 region, where it
reaches approximately 20 magnitudes \citep{lombardi06}. We selected cores 
located in all three
regions of the Pipe ({\it bowl}, {\it stem}, and B59).
	 \label{fig:pipe}
    }  \end{figure*}

The Pipe Nebula is a massive ($10^4$~$M_{\sun}$: \citealp{onishi99}) filamentary ($\sim$15~pc long and $\sim$3~pc wide)
dark cloud located in the southern sky $\sim$5\degr\ apart from the Galactic Center. Its short distance to the Sun (145
pc: \citealp{alves07}) places this complex in the group of nearby molecular clouds which serve as good laboratories for
star formation surveys.  Despite the large reservoir of mass, the Pipe Nebula molecular cloud is characterized by being
apparently quiescent, with a very low star-formation efficiency ($\sim$0.06\% for the entire cloud,
\citealp{forbrich09}).  Barnard 59 (B59), located at the northwestern end of the cloud, has formed a small cluster of
low-mass stars \citep{brooke07}.  The low global star forming efficiency of the cloud contrasts with that of other nearby
molecular clouds such as Ophiuchus or Taurus, where an important star-formation activity is observed. The Pipe Nebula is,
hence, an excellent place to study the initial  conditions of star formation at scales of a few parsecs. 

The first extensive survey toward the Pipe Nebula was done by \cite{onishi99} through single dish observations of CO
isotopologues. These authors were the first to suggest a clumpy distribution for the dense gas by detecting compact
C$^{18}$O cores in the main body of the cloud. It was not until the last few years that several surveys
\citep{lombardi06,muench07,brooke07,rathborne08,lada08} were carried out to explore the physical properties of the cloud.
\cite{lombardi06} use 2MASS data to construct a high resolution extinction map of the Pipe Nebula through which they
identify a large number of high extinction cores with typical masses between 0.2 and 5~$M_{\odot}$. Molecular line
observations reveal that they are starless cores in a very early evolutionary stage, associated with dense
($10^4$~cm$^{-3}$), relatively cold ($9.5 \leq T_K \leq 17$~K), and fairly quiescent gas (typical line widths of
0.4~\kms, \citealp{muench07,rathborne08}). Non-thermal gas motions inside the cores are sub-sonic and mass independent.
Therefore, thermal pressure appears to be the dominant source of internal pressure. In addition, these cores appear to be
pressure confined, but gravitationally unbound \citep{lada08}. 

Recently, \citet{alves08} performed an optical polarimetric survey toward the diffuse gas in the Pipe Nebula.  They find
a large scale magnetic field that appears to be mostly perpendicular  to the cloud main axis. The magnetic field exerts a
pressure ($\sim$10$^6$~K\,cm$^{-3}$) that is likely responsible for driving the collapse of the  gas and dust cloud along
the field lines. The polarization properties significantly change along the Pipe Nebula. This fact allowed the authors to
distinguish three regions in the cloud: B59, the \textit{stem}, and the \textit{bowl} (see Fig.~\ref{fig:pipe}).   B59
shows low polarization levels but high dispersion of the polarization position angles. Moving through the \textit{stem}
toward the \textit{bowl}, the polarization level increases and the dispersion decreases.  These authors propose that
these three regions might be in different evolutionary stages. B59 is the only magnetically supercritical region and  the
most evolved of the Pipe Nebula, the \textit{stem} would be at an earlier evolutionary stage, with material still
collapsing, and finally, the \textit{bowl} would be at the earliest stage, with cloud fragmentation just started.


\begin{table}[tbp]
\caption{
Source list.
}
\begin{tabular}{ccccc}
\hline\hline
& \multicolumn{1}{c}{$\alpha$(J2000)} 
& \multicolumn{1}{c}{$\delta$(J2000)} 
& \multicolumn{1}{c}{$\texttt{v}_{\rm LSR}$}
& \multicolumn{1}{c}{} \\
\multicolumn{1}{c}{Source$^a$} &
\multicolumn{1}{c}{h m s} &
\multicolumn{1}{c}{$\degr$ $\arcmin$ $\arcsec$} &
\multicolumn{1}{c}{(km s$^{-1}$)} &
\multicolumn{1}{c}{Region$^b$} \\
\hline
Core 14  & 17 12 34.0 & -27 21 16.2 & +3.6 & B59\\
Core 40  & 17 21 16.4 & -26 52 56.7 & +3.3 & {\it Stem}\\
Core 48  & 17 25 59.0 & -26 44 11.8 & +3.6 & {\it Stem}\\
Core 109 & 17 35 48.5 & -25 33 05.8 & +5.8 & {\it Bowl}\\ 
\hline
\end{tabular}

(a) According to \citet{lombardi06} numbering. \\
(b) According to \citet{alves08} diffuse gas polarimetric properties.
\label{tab_source}
\end{table}


Based on \cite{alves08} results, we selected a sample of cores distributed in the different regions of the Pipe Nebula.
We started an extensive molecular survey of these cores using the IRAM~30-m telescope. The aim of this study is to probe
their chemical evolutionary stage, which could be related with the dynamical age according to chemical modeling of
starless cores \citep{taylor96,morata03,tafalla06}.  These models predict that some molecules, such as carbon-containing
molecules, are formed very early in the chemical evolution, and are known as early-time molecules. These species are
expected to be abundant in chemically young or low density cores, and most of them seem to be affected earlier by
depletion effects (see e.g., \citealp{taylor96,ohashi99,bergin01,tafalla06}).  Other species, such as nitrogen-bearing
molecules and deuterated species, require a longer time to form. Thus, they are formed later in the chemical evolution
and are known as late-time molecules. They are not expected to be depleted until densities of 10$^6$~cm$^{-3}$ are
reached (see e.g. \citealp{caselli02,flower06,bergin07,aikawa08}).  The qualitative comparison of the relative abundances
of different types of molecules in each core can provide us with some clues about their possible evolutionary stage. 
From the observational point of view, there have been several authors that have studied the evolutionary stage of pre-
and protostellar cores through molecular surveys.  For instance, \citet{kontinen00} have used a large sample of molecules
in a prestellar and a  protostellar core. They find very different chemical compositions, specially in N$_2$H$^+$  and
long carbon-chain molecule abundances. The former is typical of a pure gas-phase chemistry, while the latter require an
evolved chemistry to form. According to time-dependent chemistry models they interpret the differences as  different
stages of chemical evolution. Later, \citet{tafalla04} made a chemical analysis of L1521E, which helped to determine the
extreme youth of this prestellar core.  From the theoretical point of view, \citet{aikawa03} have simulated the evolution
of a prestellar core and identified the different molecular abundances at different evolutionary stages to finally
compare the results with the sample of \citet{tafalla02}. \citet{morata03,morata05} have used the modeling results of
\citet{taylor96} to compare with observations toward the L673 molecular cloud. 

Based on this, we observed a set of early- and late-time molecules (see Table~\ref{tab_obs}) toward the selected cores
for a subsequent comparison. In addition, we mapped the 1.2~mm dust continuum emission of the cores to obtain  a complete
description of the structure, chemistry and evolutionary stage of the four selected Pipe Nebula cores.

\section{Observations and data reduction\label{obs}}

\subsection{MAMBO-II observations\label{obs_mambo}}

We mapped the Cores~14, 40, 48, and 109 (according to the \citealp{lombardi06} numbering) at 1.2~mm with the 117-receiver
MAMBO-II bolometer (array diameter of 240$''$) of the 30-m IRAM telescope in Granada (Spain). The positions and velocity
of the local standard of rest ($\texttt{v}_{\rm LSR}$) for each core are listed in Table~\ref{tab_source}. The
observations were carried out in April and May 2009 and in January and March 2010, in the framework of a flexible
observing pool. A total of 13 usable maps were selected for analysis: 3 for Cores 14, 40, and 109, and 4 for Core 48. The
weather conditions were good, with zenith optical depths between 0.1 and 0.3 for most of the time. The maps were taken at
an elevation of $\la$25$^{\circ}$ due to the  declination of the sources.

The beam size of the telescope is $\sim$11$''$ at the effective frequency of 250~GHz. The sources were observed with the
on-the-fly technique, with the secondary chopping between 46$''$ and 72$''$ parallel to the scanning direction of the
telescope. The telescope was constantly scanning at a speed of 8$''$ s$^{-1}$ for up to 65~s. This resulted in typical
integration times for each map of $\sim$1 hour. When possible, each source was mapped with different scanning directions
(in equatorial coordinates) or rotating the secondary mirror of the telescope to avoid scanning artifacts in the final
maps.  We measured the zenith optical depth with a skydip and checked pointing and focus before and after each map.  The
average corrections for pointing and focus stayed below 3$''$ and 0.2~mm, respectively.  Flux density calibrators were
observed every few hours.

The data were reduced using MOPSIC with the iterative reduction strategy developed by \citet{kauffmann08}. The main
advantages of the new scheme are that ({\it i}) sources much larger than in the classical approach can be recovered,
({\it ii}) the signal-to-noise ratio (SNR) of the final map increases, and ({\it iii}) they suffer from less artifacts.
The figures were created using the GREG package, from the GILDAS\footnote{MOPSIC and GILDAS data reduction packages are
available at http://www.iram.fr/IRAMFR/GILDAS}  software.

All the maps have been convolved with a 21.$\!''$5 Gaussian, larger than the telescope beam, in order to improve the SNR,
and to smooth the appearance of the maps. The size of the Gaussian was chosen to be the one of the CN~(1--0) molecular
transition (see Table~\ref{tab_obs}) which provides good spatial resolution and large SNR for the four maps.

\subsection{Line observations\label{obs_line}}

\begin{figure*}[htpb]
   \centering
   \includegraphics[width=12cm,angle=-90]{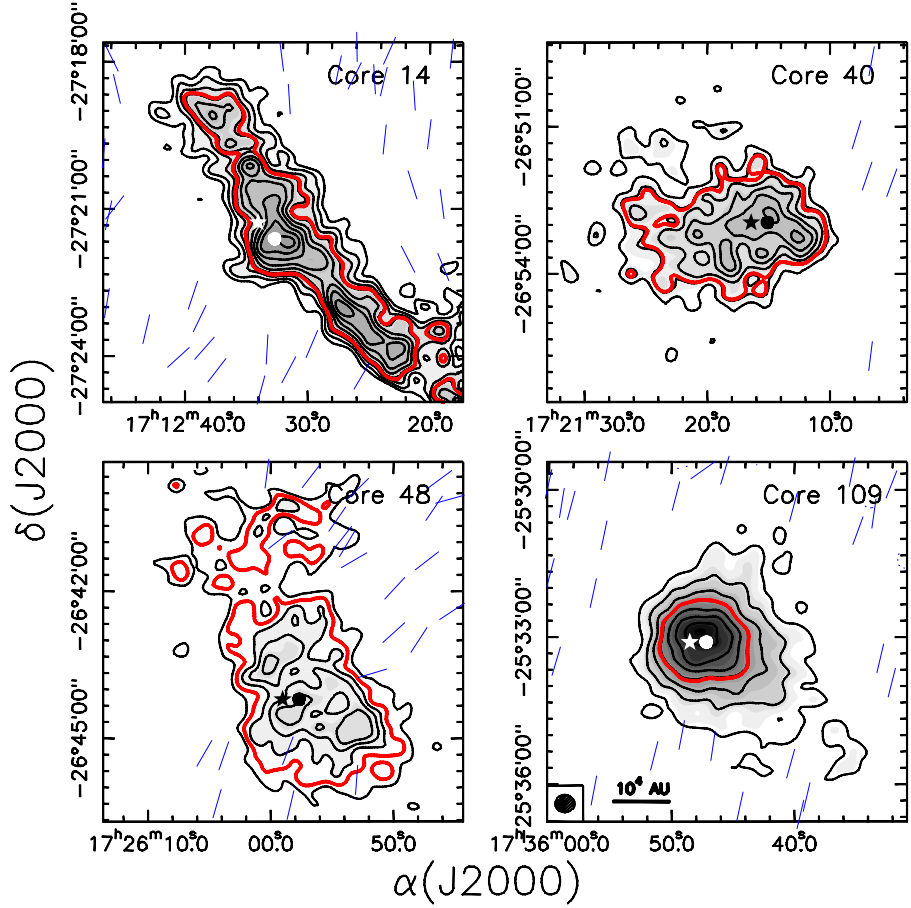}
      \caption{IRAM 30-m MAMBO-II maps of the dust continuum emission at 1.2~mm
      toward four cores of the Pipe Nebula.  The greyscale levels for all the maps are
      3 to 18 times 5.75~mJy\,beam$^{-1}$.
      The contour levels are 3 to 11 times
      $\sigma$  in steps of 1-$\sigma$, for Cores 14, 40 and 48, 
      and 3 to 21 $\sigma$ in steps of 3-$\sigma$ for Core~109. 1-$\sigma$ is 
      4.5, 5.0, 3.5, and 4.5~mJy\,beam$^{-1}$ for Core~14, 40, 48,
      and 109, respectively. 
      The red thick contour mark the half maximum emission level of the source (see Table~\ref{tab_param_mambo}).
      Black or white stars indicate the C$^{18}$O pointing center reported 
      by \cite{muench07}, which is
      very close to the  visual extinction peak position of each core 
      \citep{lombardi06}. 
      Black or white filled circles indicate the position 
      where line observations have been performed, 
      close to the dust continuum emission maximum which falls into the beam area.
      The blue vectors depict the
      magnetic field vector found by \cite{franco10}. 
      Note that for Core~40 there are no optical polarimetry measurements
      on the east side due to the high visual extinction.
      In the bottom left corner of the bottom right panel the
      convolved beam and the spatial scale for the maps is shown.
	 \label{fig:mambo}
              }
   \end{figure*}

We made several pointed observations within the regions of the Cores~14, 40, 48, and 109  with the heterodyne receivers
of the 30-m IRAM telescope (ABCD and EMIR receivers). The observations were carried out in three epochs. The first epoch
was August and September 2008.  We used the capability of the telescope to perform simultaneous observations at different
frequencies to observe the emission of the  C$_3$H$_2$~(2$_{\rm 1,2}$--1$_{\rm 1,0}$), HCN~(1--0), N$_2$H$+$~(1--0),
CS~(2--1), CN~(1--0), N$_2$D$+$~(2--1), DCO$^+$~(3--2), CN~(2--1),  N$_2$D$+$~(3--2) and H$^{13}$CO$^+$~(3--2) molecular
transitions arranged in 3 different frequency setups covering the 3, 2, 1.3 and 1.1~mm bands. To do this, we combined the
A100/B100/A230/B230 and A100/D150/A230/D270 SIS heterodyne receivers. The observational strategy was first to observe
several positions with a 20$''$ spacing centered on the C$^{18}$O pointing center reported by \cite{muench07} (depicted
by star symbols in Fig.~\ref{fig:mambo}), which is very close to the  visual extinction peak position of each core
\citep{lombardi06}. The visual extinction peak is assumed to be the densest region of the core, and it was defined as
core center by \cite{muench07}. The second and third epochs were August 2009 and June 2010, respectively, both using the
new EMIR E0/E1/E2 receivers. We observed deeper toward the position of the grid of first epoch closer to the dust
continuum peak (see circle symbols in Fig.~\ref{fig:mambo}). We observed also the C$^{34}$S~(2--1) molecular transition.
Table~\ref{tab_obs} shows the transitions and frequencies observed. We used the VESPA autocorrelator as the spectral
backend, selecting a channel resolution of 20~kHz, which provided a total bandwidth of 40~MHz.  The corresponding
velocity resolutions, main-beam efficiencies and half-power beam widths at all the observed frequencies are also listed
in Table~\ref{tab_obs}. We used the frequency-switching mode  with a frequency throw between 3.83 and 22.98~MHz,
depending on the transition. System temperatures  in nights considered ``good'' were between 200 to 275~K at 3~mm, and
between 440 and 960~K at 1~mm ($T_{\rm sys}$ reached 450~K and 3200~K in bad nights, respectively). Pointing was checked
every 2 hours.


\begin{table}[t]
\caption{
Molecular transitions observed toward the Pipe Nebula cores with the IRAM 30-m antenna.
}
\scriptsize
\begin{tabular}{cc r @{.} l r @{.} l r @{/} l cc}
\hline\hline
&&\multicolumn{2}{c}{Frequency} 
&\multicolumn{2}{c}{Beam} 
&\multicolumn{2}{c}{Beam} 
&\multicolumn{1}{c}{$\Delta {\tt v}$ $^b$} \\
 \multicolumn{1}{c}{Molecule} 
&\multicolumn{1}{c}{Transition} 
&\multicolumn{2}{c}{(GHz)} 
&\multicolumn{2}{c}{($''$)} 
&\multicolumn{2}{c}{efficiency$^a$} 
&\multicolumn{1}{c}{(km\,s$^{-1}$)}
&\multicolumn{1}{c}{Type$^c$}\\
\hline
C$_3$H$_2$      &(2$_{1,2}$--1$_{1,0}$)  & 85  & 3389 & 29 & 0  &0.78&0.81 &0.07  & E\\
HCN             &(1--0)                  & 88  & 6318 & 28 & 0  &0.78&- &0.07  & E\\
N$_2$H$^+$      &(1--0)                  & 93  & 1762 & 26 & 5  &0.77&0.81 &0.06  & L\\
C$^{34}$S	&(2--1)			 & 96  & 4130 & 26 & 0  &-&0.81 &0.06  & E\\
CS              &(2--1)                  & 97  & 9809 & 25 & 5  &0.76&0.81 &0.06  & E\\
CN              &(1--0)                  & 113 & 4909 & 21 & 5 &0.75&0.81 &0.05  & E\\
N$_2$D$^+$      &(2--1)                  & 154 & 2170 & 15 & 0  &0.77&0.74 &0.04  & L\\
DCO$^+$         &(3--2)                  & 216 & 1126 & 10 & 5  &0.57&0.63 &0.03  & L\\
CN              &(2--1)                  & 226 & 8747 & 10 & 0  &0.53&0.63 &0.03  & E\\
N$_2$D$^+$      &(3--2)                  & 231 & 3216 & 10 & 0  &0.67&0.63 &0.03  & L\\
H$^{13}$CO$^+$  &(3--2)                  & 260 & 2554 &  9 & 0  &0.53&0.63 &0.02  & L\\
\hline
\end{tabular}
(a) ABCD and EMIR receiver, respectively\\
(b) Spectral resolution.\\
(c) E = Early-time. L = Late-time. See \S\ref{intro} and \S\ref{dis_chem} for details.
\label{tab_obs}
\end{table}

We reduced the data using the
CLASS package of the GILDAS$^1$ software. We obtained the line parameters
either from a Gaussian fit or from calculating their statistical moments when the profile was not
Gaussian.

\begin{table*}[htpb]

\caption{
1.2~mm continuum emission parameters.
}
\begin{tabular}{ccccccccccc}
\hline
&
\multicolumn{1}{c}{$\alpha$(J2000)$~^a$} &
\multicolumn{1}{c}{$\delta$(J2000)$~^a$} &
\multicolumn{1}{c}{$T_{\rm dust}$} &
\multicolumn{1}{c}{RMS} &
\multicolumn{1}{c}{$S_{\nu}$} &
\multicolumn{1}{c}{$I_\nu^{\textrm{\tiny{Peak}}}$} &
\multicolumn{1}{c}{Diameter} &
\multicolumn{1}{c}{$N_{\rm H_2}$ $^b$} &
\multicolumn{1}{c}{$n_{\rm H_2}$ $^b$} &
\multicolumn{1}{c}{Mass $^b$} 
 \\
\multicolumn{1}{c}{Source} &
\multicolumn{1}{c}{h m s} &
\multicolumn{1}{c}{$\degr$ $\arcmin$ $\arcsec$} &
\multicolumn{1}{c}{(K)} &
\multicolumn{1}{c}{(mJy beam$^{-1}$)} &
\multicolumn{1}{c}{(Jy)} &
\multicolumn{1}{c}{(mJy  beam$^{-1}$)} &
\multicolumn{1}{c}{(pc)} &
\multicolumn{1}{c}{(10$^{21}$cm$^{-2}$)} &
\multicolumn{1}{c}{(10$^{4}$cm$^{-3}$)} &
\multicolumn{1}{c}{($M_{\odot}$)} \\
\hline

Core 14	(filament)
	&\multirow{2}{*}{17 12 31.5} &\multirow{2}{*}{$-$27 21 41.0} & \multirow{2}{*}{12.0$^c$}
	& 	\multirow{2}{*}{4.5}	
	& 2.56  
	&\multirow{2}{*}{51.6}
	&	0.106	&	12.21	&	5.59	&	2.87	\\

Core 14	(core)
	&			&			&	
	&			
	&	1.24 
	&	
	&	0.071	&	13.27	&	9.09	&	1.40		\\

Core 40	
	&	17 21 14.7&	$-$26 52 47.8	&	10.3$^c$ 
	&	5.0	
	&	1.73
	&	42.0		
	&	0.104	&	11.05      &	5.16	&	2.51		\\

Core 48 
	&	17 25 57.3&	$-$26 44 22.3	&	10.0$^d$   
	&	3.5	
	&	1.44
	&	27.9		
	&	0.127	&	6.14$^d$	&	2.35$^d$&	2.09$^d$	\\

Core 109
	&	17 35 47.7&	$-$25 32 52.9	&	9.5$^c$  
	&	4.5	
	&	2.76 
	&	105.3
	&	0.063	&	47.60	&	36.57	&	4.00		\\
\hline
\end{tabular}

(a) Pointing position of the chemical observations
which lies inside the same beam area of the dust continuum emission peak.\\
(b) Assuming $\kappa_{250~GHz}$$=$0.0066~cm$^2$ g$^{-1}$ as a medium value 
between dust grains with thin
and thick ice mantles \citep{ossenkopf94}. See Appendix~\ref{app_dust}
 for details on the calculation. \\
(c) Adopted to be equal to the kinetic temperature derived for NH$_3$ \citep{rathborne08}.\\
(d) No kinetic temperature estimate, therefore we assumed 10~K 
based on the temperatures of the other cores
\citep{rathborne08}. \\ 
\label{tab_param_mambo}
\end{table*}


\section{Results and analysis}


\subsection{Dust continuum emission\label{res_dust}}

In Fig.~\ref{fig:mambo} we present the MAMBO-II maps of the dust continuum emission at 1.2~mm toward the four selected
cores of the Pipe Nebula, convolved to a 21.$\!''$5 beam. Table~\ref{tab_param_mambo} gives the peak position of the
1.2~mm emission after convolution with a Gaussian, the dust temperature \citep{rathborne08}, the RMS noise of the
emission, the flux density and the value of the emission peak. Additionally, we also give the derived FWHM equivalent
diameter, which is the diameter of the circular area equal to the area within the FWHM level, depicted by a red contour
in Fig.~\ref{fig:mambo}. Table~\ref{tab_param_mambo} also lists the H$_2$ column and volume density, as well as the mass
for each core. These parameters are derived from the emission within the 3-$\sigma$ level and discussed in
$\S$\ref{sec_discuss}.

The flux density of the cores ranges between $\sim$1.24 and $\sim$2.76~Jy. Note, however, that the extinction maps show
that the studied cores are surrounded by a diffuse medium (see Fig.~\ref{fig:pipe} and \citealp{lombardi06}).  The
on-the-fly reduction algorithms assume that the map limits  have a zero emission level. Due to the presence of the
diffuse material, this could not be true for the observed cores, and, therefore, the  measured flux density of the maps
might be lower than the actual value. We derived average H$_2$ column densities ($N_{\rm H_2}$, see
Appendix~\ref{app_dust})  toward the dust continuum emission peak for the different resolutions (listed in
Table~\ref{tab_dust_col_dens}) of the detected molecular transitions (see Table~\ref{tab_detect}). We derived their
abundances with respect to H$_2$. The results are discussed in $\S$~\ref{sec_discuss}. 

The maps of Fig.~\ref{fig:mambo} show the different morphology of the cores. Following the results of \citet{alves08}, it
is interesting to compare the shape of the cores with their location along the Pipe Nebula. Core~14, located in B59,
belongs to a clumpy and filamentary structure of $\sim$500$''$ ($\sim$0.35~pc) elongated along the NE-SW direction. This
is in perfect agreement with previous extinction maps \citep{lombardi06,roman09}. On the other hand, Core~109, located in
the {\it bowl}, shows a compact and  circular morphology with a FWHM of $\sim$90$''$ ($\sim$0.063~pc). Cores~40 and 48,
both located in the {\it stem}, have elliptical morphologies with extended diffuse emission.

   \begin{figure}
   \centering
   \includegraphics[width=9cm,angle=0]{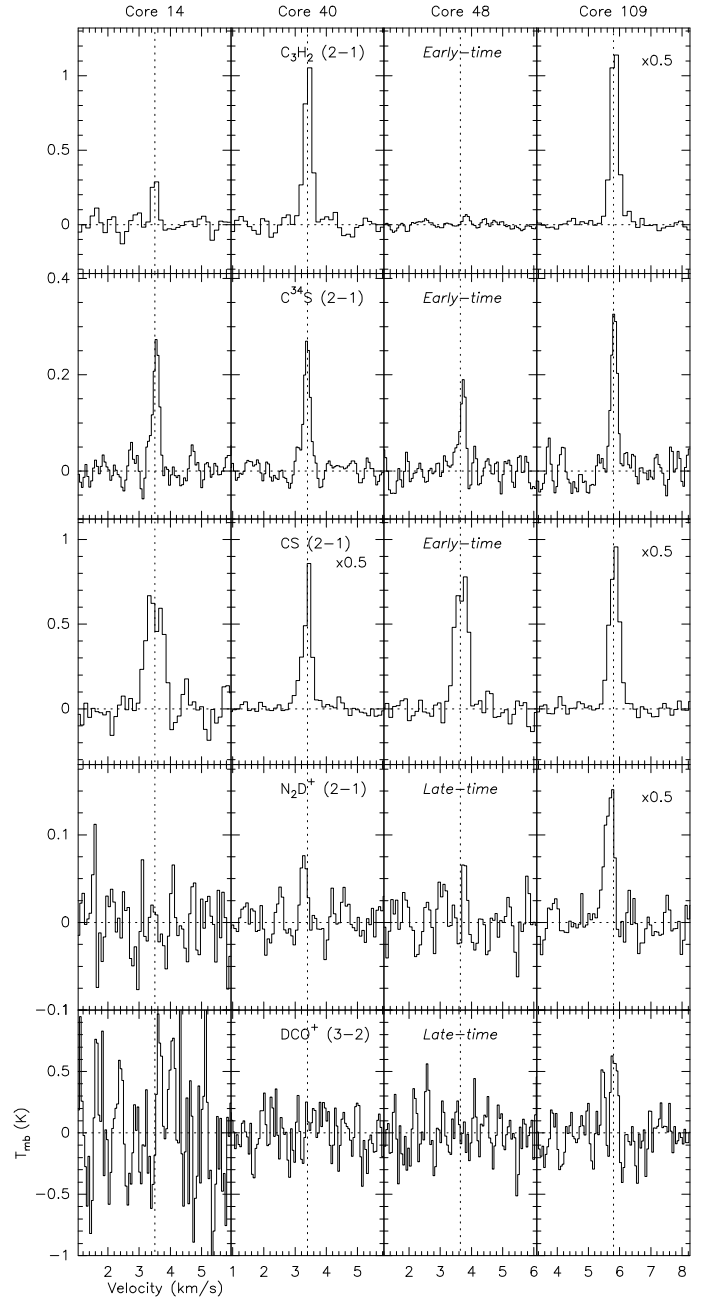}
      \caption{  
      IRAM 30-m line spectra of the molecular transitions with no hyperfine components
      toward the four selected cores of the Pipe Nebula
      (see Table~\ref{tab_source}). The name of the core is indicated above the top
      panel of each column.  Rows correspond to a single molecular
      transition specified on the second column. The velocity range is 5~km~s$^{-1}$,
      and it is
      centered on the $\texttt{v}_{\rm LSR}$ of each core is marked with a vertical dotted line.
      Vertical axis show the $T_{\rm MB}$ of the emission, and the zero level is marked by a
      horizontal dotted line. 
      Some of the Core~109 spectra, with the highest measured $T_{\rm MB}$,
      have been divided by 2 to fit to the common scale.
	 \label{fig:lines1}
              }
   \end{figure}

   \begin{figure*}
   \centering
   \includegraphics[width=11cm,angle=-90]{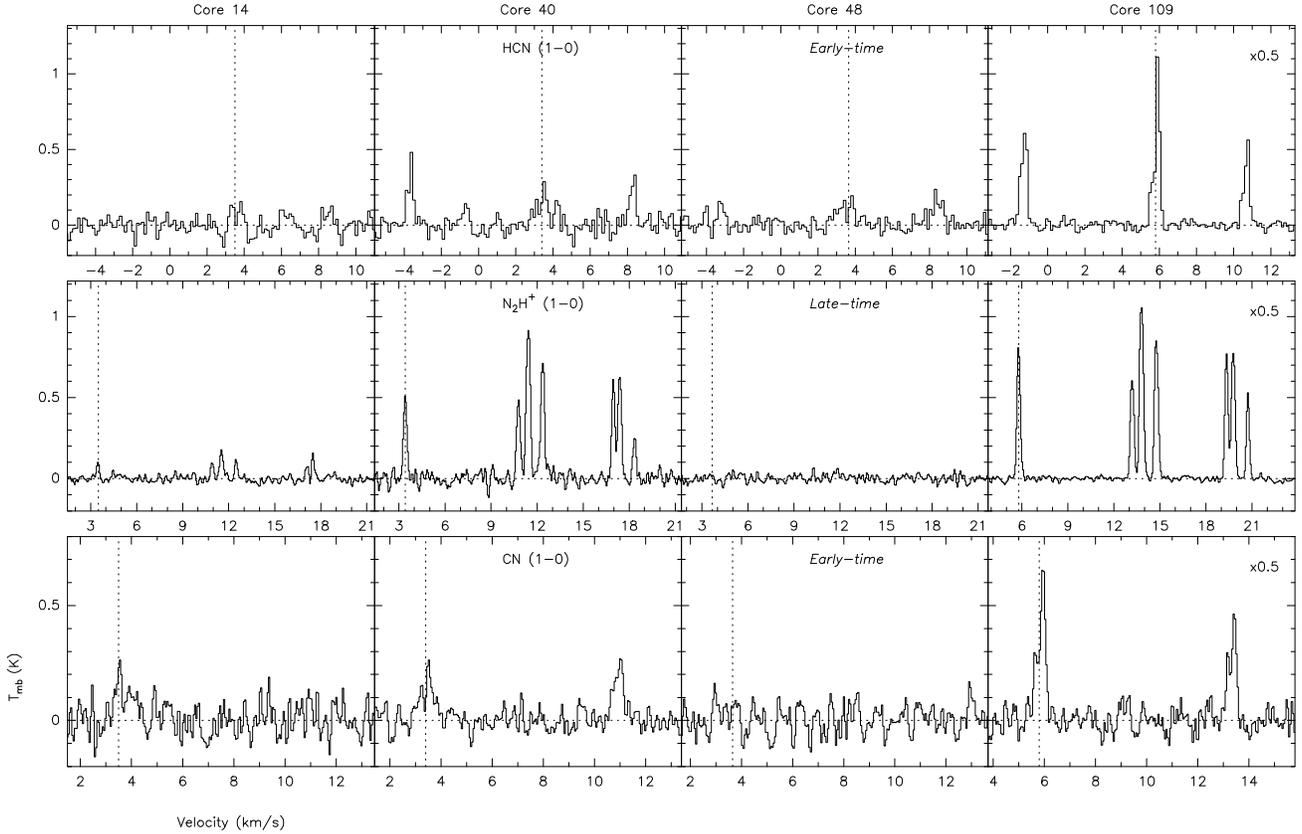}
      \caption{  
      IRAM 30-m line spectra of the molecular transitions with hyperfine components
      toward the four selected cores of the Pipe Nebula
      (see Table~\ref{tab_source}). The name of the core is indicated above the top
      panel of each column.  Rows correspond to a single molecular
      transition specified on the second column.
      The velocity range is 16.5, 20 and 12~km~s$^{-1}$ for HCN~(1--0), N$_2$H$^+$~(1--0)
      and CN~(1--0) respectively.
      The $\texttt{v}_{\rm LSR}$ of each core is marked with a vertical dotted line.
      Vertical axis show the $T_{\rm MB}$ of the emission, and the zero level is marked by a
      horizontal dotted line. 
      Core~109 spectra, with the highest measured $T_{\rm MB}$,
      have been divided by 2 to fit to the common scale.
	 \label{fig:lineshfs}
              }
   \end{figure*}


\begin{table}[tbp]
\caption{
H$_2$ column densities$^a$, $N_{\rm H_2}$, of the Pipe Nebula cores in cm$^{-2}$.
}
\begin{tabular}{c cccc}
\noalign{\smallskip}
\hline\hline\noalign{\smallskip}
\multicolumn{1}{c}{Source}  &  10.$\!''$5 &  15.$\!''$0 & 21.$\!''$5 & 27.$\!''$0\\
\noalign{\smallskip}
\hline\noalign{\smallskip}
Core 14  & $1.75\times10^{22}$ & $1.38\times10^{22}$ & $1.21\times10^{22}$ & $1.11\times10^{22}$\\
Core 40  & $1.32\times10^{22}$ & $1.28\times10^{22}$ & $1.12\times10^{22}$ & $1.07\times10^{22}$\\
Core 48  & $1.09\times10^{22}$ & $8.88\times10^{21}$ & $7.38\times10^{21}$ & $6.99\times10^{21}$\\
Core 109 & $4.19\times10^{22}$ & $3.73\times10^{22}$ & $3.23\times10^{22}$ & $3.08\times10^{22}$\\
\hline
\end{tabular}

(a) Average column densities are calculated within one beam area 
toward the dust continuum emission peak.
The values of $\kappa_{250~GHz}$ and $T_{\rm dust}$ are the same as for Table~\ref{tab_param_mambo}.
These values are combined with the molecular column densities to find the molecular abundances
in the same beam area. The correspondence is:
10.$\!''$5 with DCO$^+$, 
15.$\!''$0 with N$_2$D$^+$, 
21.$\!''$5 with  CN~(1--0) and, finally, 
27.$\!''$0 with C$_3$H$_2$, HCN, N$_2$H$^+$, CS and C$^{34}$S.\\
\label{tab_dust_col_dens}
\end{table}


\subsection{Molecular survey of high density tracers\label{res_line}}


\begin{table}[hbtp]
\caption{
Summary of detections and upper levels in K toward the Pipe Nebula cores$^a$.
}
\begin{tabular}{ccccc}
\hline\hline
\multicolumn{1}{c}{Molecular}  & 
\multicolumn{4}{c}{Core}\\
\cline{2-5}
\multicolumn{1}{c}{transitions}    & 
\multicolumn{1}{c}{14} & 
\multicolumn{1}{c}{40} & 
\multicolumn{1}{c}{48} & 
\multicolumn{1}{c}{109}\\
\hline
C$_3$H$_2$~(2$_{1,2}$--1$_{1,0}$)  & $\surd$    & $\surd$ & $<$0.07 & $\surd$ \\
HCN~(1--0)  & $<$0.21  & $\surd$  & $\surd$  & $\surd$ \\
N$_2$H$^+$~(1--0) & $\surd$ & $\surd$ & $<$0.07 & $\surd$ \\
C$^{34}$S~(2--1)   & $\surd$   & $\surd$  & $\surd$   & $\surd$ \\
CS~(2--1)   & $\surd$   & $\surd$  & $\surd$   & $\surd$ \\
CN~(1--0)   &$\surd$   & $\surd$& $<$0.17 & $\surd$ \\
N$_2$D$^+$~(2--1) & $<$0.12& $\surd$ & $<$0.08&  $\surd$\\
DCO$^+$~(3--2) & $<$1.71&  $<$0.61& $<$0.76& $\surd$ \\
CN~(2--1) & $<$0.97&  $<$1.70& $<$0.76& $<$0.90 \\
N$_2$D$^+$~(3--2) & $<$1.01&  $<$0.93& $<$1.94& $<$0.91\\
H$^{13}$CO$^+$~(3--2) & $<$1.52& $<$1.40& $<$2.38&  $<$1.34\\
\hline
\end{tabular}

(a) The transitions marked with $\surd$ have been detected toward the corresponding core. Otherwise, the 
3$\sigma$ upper limit is shown.
\label{tab_detect}
\end{table}


\cite{muench07} reported C$^{18}$O pointed observations toward the Pipe Nebula cores measured with a resolution of
56$''$. As seen in their Fig.~1, the position of the C$^{18}$O is very close to the visual extinction peak position of
each core \citep{lombardi06}. Our higher resolution maps show a peak position offset for all the cores. As listed in
Tables~\ref{tab_source} and \ref{tab_param_mambo}, and as shown in Fig.~\ref{fig:mambo}, the dust continuum peak does not
coincide exactly with the $A_\texttt{v}$ peak (stars in Fig.~\ref{fig:mambo}). However, the difference  is compatible
with the angular resolution of the extinction maps. We decided to present only molecular line data of the observed
positions closer to the dust continuum emission peak (circles in Fig.~\ref{fig:mambo}) , defined as the core center and
supposed to exhibit brighter emission from molecular transitions.   The typical core size is $\sim$90$''$ or larger (see
Table~\ref{tab_param_mambo}). The beam size of the detected lines, except for N$_2$D$^+$~(2--1) and DCO$^+$~(3--2), range
from 21.$\!''$5 to 29.$\!''$0, while the initial grid of pointed position had a separation of 20.$\!''$0, thus the
emission  peak stays within the beam area for these molecular transitions. Therefore, the molecular line properties that
we obtain are representative of the chemistry of the core center.

Table~\ref{tab_detect} summarizes the detections or the 3$\sigma$ upper limits of the non detections toward each core. 
Table~\ref{tab_lines} give the parameters of the detected lines. In Figs.~\ref{fig:lines1} and \ref{fig:lineshfs} we show
the spectra of the different molecular transitions observed toward the dust continuum emission peak of each core. 
Core~109 shows the stronger emission in all the detected transitions in our sample. This is the core with the most
compact and circular morphology (see Fig.~\ref{fig:mambo}). Core~40 shows also emission in the six molecular transitions
at 3~mm (C$_3$H$_2$~(2$_{\rm 1,2}$--1$_{\rm 1,0}$), HCN~(1--0), N$_2$H$^+$~(1--0), C$^{34}$S~(2--1), CS~(2--1), and
CN~(1--0)),  although their intensities are lower than for Core~109.  Core~14 shows emission in all the 3~mm transitions
except in HCN~(1--0). Finally, Core~48 only shows emission in CS~(2--1), C$^{34}$S~(2--1), and HCN~(1--0).

\begin{table*}[htbp]
\caption{
Line parameters$^a$.
}
\begin{tabular}{c c r@{.}l r@{.}l r@{.}l r@{.}l r@{.}l r@{.}l c}
\hline\hline
\multicolumn{1}{c}{Molecular}
&
& \multicolumn{2}{c}{$T_{\rm MB}$ $^b$} 
& \multicolumn{2}{c}{$A \times \tau$ $^c$} 
& \multicolumn{2}{c}{$\int T_{\rm MB}\mbox{d}\texttt{v}$ $^b$} 
& \multicolumn{2}{c}{$\texttt{v}_{\rm LSR}$} 
& \multicolumn{2}{c}{$\Delta\texttt{v}_{\rm LSR}$} 
\\
\multicolumn{1}{c}{transition}
& \multicolumn{1}{c}{Source} 
& \multicolumn{2}{c}{(K)} 
& \multicolumn{2}{c}{(K)} 
& \multicolumn{2}{c}{(K\, km\,s$^{-1}$)} 
& \multicolumn{2}{c}{(km\,s$^{-1}$)} 
& \multicolumn{2}{c}{(km\,s$^{-1}$)}
& \multicolumn{2}{c}{$\tau$ $^d$}
& \multicolumn{1}{c}{Profile$^e$} \\
\hline

C$_3$H$_2$~(2$_{1,2}$--1$_{1,0}$) 
&Core 14 &    0&37(6) &\multicolumn{2}{c}{-}&    0&086(11) &    3&502(14) &    0&22(3) &\multicolumn{2}{c}{-}& G\\
&Core 40 &    1&19(5) &\multicolumn{2}{c}{-}&    0&347(9) &    3&420(4) &    0&273(9) &\multicolumn{2}{c}{-}& G\\
&Core 109 &    2&74(6) &\multicolumn{2}{c}{-}&    0&799(13) &    5&8340(20) &    0&274(5) &\multicolumn{2}{c}{-}& G\\
\hline

CS~(2--1)
&Core 14 &    0&69(10) &\multicolumn{2}{c}{-}&    0&41(3) &    3&439(21) &    0&45(4) &		10&8(1.1)	& SA\\
&Core 40 &    1&94(7) &\multicolumn{2}{c}{-}&    0&560(17) &    3&369(4) &    0&415(14) &	3&1(3)	& NS\\
&Core 48 &    0&79(7) &\multicolumn{2}{c}{-}&    0&402(18) &    3&684(11) &    0&477(22) &      6&0(6)       & SA\\
&Core 109 &    1&93(8) &\multicolumn{2}{c}{-}&    0&743(17) &    5&836(4) &    0&361(9) &	4&2(4)	& G\\
\hline

C$^{34}$S~(2--1)
&Core 14 &    0&267(25) & \multicolumn{2}{c}{-}&    0&068(5) &    3&545(8) &    0&241(20) &  0&5(1)& G\\
&Core 40 &    0&268(16) &  \multicolumn{2}{c}{-}&   0&069(3) &    3&381(5) &    0&241(13) &  0&14(1)& G\\
&Core 48 &    0&187(23) & \multicolumn{2}{c}{-}&    0&041(4) &    3&729(11) &    0&20(3) &   0&26(3)& G\\
&Core 109 &    0&34(3) &  \multicolumn{2}{c}{-}&   0&083(5) &    5&825(7) &    0&233(17) &   0&19(2)& G\\
\hline

N$_2$D$^+$~(2--1)~$^f$
&\multirow{1}{*}{Core 40}  &   0&084(20) & \multicolumn{2}{c}{-}&    0&019(3) &    3&280(15) &    0&21(3)   & \multicolumn{2}{c}{-}& G\\
&\multirow{1}{*}{Core 109} &   0&31(4)   & \multicolumn{2}{c}{-}&    0&109(7) &    5&673(11) &    0&331(22) & \multicolumn{2}{c}{-}& G\\
\hline

DCO$^+$~(3--2)
& Core 109 &    0&70(11) &\multicolumn{2}{c}{-}&    0&151(18) &    5&828(13) &    0&202(21) &\multicolumn{2}{c}{-}& G\\
\hline

HCN~(1--0)
& \multirow{1}{*}{Core 40} &\multicolumn{2}{c}{-}&    1&55(11) &\multicolumn{2}{c}{-}&    3&410(16) &    0&334(22) &    6&0(5) & NS\\
& \multirow{1}{*}{Core 48} &\multicolumn{2}{c}{-}&    0&33(10) &\multicolumn{2}{c}{-}&    3&54(5) &    0&90(11) &    2&4(1.2) & G\\
& \multirow{1}{*}{Core 109 (1)} &\multicolumn{2}{c}{-}&    2&53(3) &\multicolumn{2}{c}{-}&    5&93(7) &    0&16(22) &    0&25(10) & NS\\
& \multirow{1}{*}{Core 109 (2)} &\multicolumn{2}{c}{-}&    6&10(3) &\multicolumn{2}{c}{-}&    5&72(7) &    0&25(22) &   10&20(10) & NS\\
\hline

N$_2$H$^+$~(1--0)
&\multirow{1}{*}{Core 14}  &\multicolumn{2}{c}{-}&    0&0341(16) &\multicolumn{2}{c}{-}&   11&500(5) &    0&206(10) &    0&10(9)  & G\\
&\multirow{1}{*}{Core 40}  &\multicolumn{2}{c}{-}&    0&219(12)  &\multicolumn{2}{c}{-}&   11&4000(19) &    0&249(5) &    0&171(25)  & G\\
&\multirow{1}{*}{Core 109} &\multicolumn{2}{c}{-}&    0&904(14)  &\multicolumn{2}{c}{-}&   13&8000(5) &    0&2150(11) &    0&467(11)  & G\\
\hline

CN~(1--0)
&\multirow{1}{*}{Core 14}       &\multicolumn{2}{c}{-}&    0&051(9) &\multicolumn{2}{c}{-}&    3&64(8)   &    0&81(15) &    0&1(7) & G\\
&\multirow{1}{*}{Core 40}       &\multicolumn{2}{c}{-}&    0&65(22) &\multicolumn{2}{c}{-}&    3&430(21) &    0&36(5)  &    3&9(1.3) & G\\
&\multirow{1}{*}{Core 109 (1)}  &\multicolumn{2}{c}{-}&    1&41(22)       &\multicolumn{2}{c}{-}&    5&930(5) &    0&162(11) &   1&13(23) & G\\
&\multirow{1}{*}{Core 109 (2)}  &\multicolumn{2}{c}{-}&    2&3(1.3)       &\multicolumn{2}{c}{-}&    5&670(7) &    0&101(16) &   4&(3) & G\\

\hline

\end{tabular}
\label{tab_lines}

(a) Line parameters of the detected lines. The former five molecular transitions have no hyperfine components.
The parameters for the transitions labeled as G (see last column) have been derived from a Gaussian fit while 
line parameters of NS and SA profiles have been derived from the intensity peak ($T_{\rm MB}$), and zero 
(integrated intensity), first (line velocity) and second (line width) order moments of the emission. 
The latter three molecular transitions have hyperfine components. The parameters have been derived using 
the hyperfine component fitting method of the CLASS package.
The value in parenthesis shows the uncertainty of the last digit/s. If the two first 
significative digits of the error are smaller than 25, two digits are given to better constrain it.\\
(b) Only for molecular transitions with no hyperfine components.\\
(c) Only for molecular transitions with hyperfine components.\\
(d) Derived from a CLASS hyperfine fit for molecular transitions with hyperfine components.
Derived numerically for CS and C$^{34}$S using Eq.~\ref{eq_num_tau}. A value of 0.3 is assumed when
no measurement is available.\\
(e) G: Gaussian profile. NS: Non-symmetric profile. SA: Self-absorption profile.\\
(f) Only the main component is detected.

\end{table*}

In addition to the line parameters, we derived the molecular column densities for all the detected species (see
Appendix~\ref{app_line} for details) which are listed in Table~\ref{tab_mol_col_dens}. For the transitions with detected
hyperfine components (HCN, N$_2$H$^+$, and CN) we derived the opacity using the hyperfine components fitting method of
the CLASS package. For the CS and C$^{34}$S molecular transitions we derived numerically the opacity using 

\begin{equation}
\frac{T_{\rm MB}({\rm C^{34}S})}{T_{\rm MB}({\rm CS})}=
\frac{1-\exp(-\tau)}{1-\exp(-\tau\ r)}\ ,
\label{eq_num_tau}
\end{equation}

\noindent where $r$ is the CS to C$^{34}$S abundance ratio, assumed to be equal to the terrestrial value (22.5,
\citealp{kim03}). We found a high opacity toward Cores~14 and 48 for CS~(2--1), 10.8 and 6.0 respectively, whose spectra
show self-absorption (see Fig.~\ref{fig:lines1}). For Cores~40 and 109 we found lower opacities,  $\tau$=3.1 and 4.2 for
CS~(2--1) respectively. We assumed optically thin emission in C$_3$H$_2$, DCO$^+$ , and N$_2$D$^+$~(2--1), the latter
with only the main hyperfine component detected. This conservative assumption could not be true, so the column densities
should be taken as lower limits.  We also derived the molecular abundances with respect to H$_2$ (see
Table~\ref{tab_abun}), taking into account the resolution for each molecular transition (see
Table~\ref{tab_dust_col_dens}).


\section{Discussion}\label{sec_discuss}

We observed four selected cores located in the different regions of the Pipe Nebula ({\it bowl},  {\it stem}, and B59),
in different molecular tracers and in dust continuum emission, to study and compare their physical and chemical
properties. The cores were selected based on the results of the optical polarimetric survey carried out by 
\cite{alves08}. In the following subsections, we discuss and compare the properties of each individual core, as well as
an overall analysis of such properties, and we try to relate our results with previous works. In particular, in the next
subsection we compare the dust continuum emission with the visual extinction maps of \citet{rathborne08} and the trend
found for the diffuse gas by \cite{alves08}.

\subsection{Comparison of visual extinction and 1.2~mm continuum emission maps\label{dis_comp}}

The beam size of our observations is 11$''$, convolved to a Gaussian of 21.$\!''$5 in  the maps shown (see
$\S$\ref{res_dust}), while that of \cite{rathborne08} is $\sim$60$''$. Hence,  our maps suffer from less beam dilution
and we can resolve smaller structures. The sensitivity limit of the $A_\texttt{v}$ observations is fixed at 1.2~mag
\citep{lada08}, which corresponds  to a column density of $\sim$1$\times10^{21}$~cm$^{-2}$ \citep{wagenblast89}.  A
conservative estimation of the sensitivity limit of our maps, in the same conditions, can be derived using the 3-$\sigma$
emission level of the noisiest continuum map convolved to a 60$''$ Gaussian. The resulting beam averaged column density, 
for a $T_K$ of 10~K, is $\sim$4$\times$10$^{20}$~cm$^{-2}$.  Therefore, as seen from the minimum column densities in the
same conditions,  our dataset have slightly better sensitivity. \cite{lada08} defines the equivalent radius of the core
using the region with emission  brighter than 3-$\sigma$, while we use the region with emission brighter than half of the
peak value. This difference prevents a direct comparison of the radii and densities. The core masses,  however, depend
only on the integrated flux density and can be compared. Our masses are on average $\sim$3.4 times smaller, ranging from
$\sim$0.9 for Core~109 to $\sim$7 for Core~14.

   \begin{figure}
   \centering
   \includegraphics[width=9cm,angle=0]{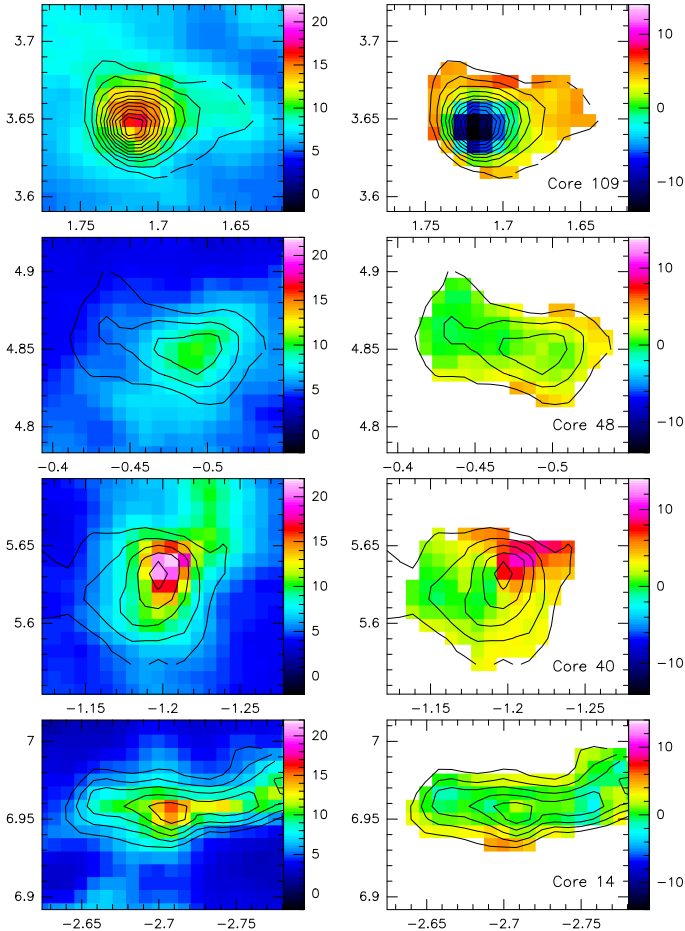}
      \caption{
      \textit{Left panels}: Color image of the visual extinction map derived from the near-IR 
observations \citep{lombardi06} superposed with the contour map of
the extinction map derived from our 1.2~mm dust continuum maps in galactic coordinates. Contours are
from 2.5 to 30 visual magnitudes by steps of 2.5. 
\textit{Right panels}: Color image of the difference between the visual extinction map
derived from the near-IR and the 1.2~mm observations within the region of the $A_\texttt{v}$ converted
 dust continuum maps with $A_\texttt{v}$$>$2.5~magnitudes.
Contours are the same as in the left panels. Core number is indicated in the lower right corner
of the panels. The color scale (in visual magnitudes) are shown in the right side of
the panels.}
       \label{fig:ext-emis}
   \end{figure}

We estimated the difference between the 2MASS extinction maps \citep{lombardi06} and the 1.2~mm dust maps. To do this we
first transformed the original near-IR extinction maps to visual extinction maps using $A_\texttt{v}=A_\texttt{k}/0.118$
\citep{dutra02}. Then, we convolved the 1.2~mm dust maps with a Gaussian of 60$''$ to have the same resolution. We
transformed the 1.2~mm dust maps to column density maps (see Appendix~\ref{app_dust} for details). We assumed a uniform
typical temperature of 10~K for all the cores. To estimate the uncertainty caused by this assumption, we also made the
calculations for temperatures of 8 and 12~K, which resulted in an average maximum variation over the whole map of
$\sim$2.4 extinction magnitudes. We also assumed for all the cores $\kappa_{250~GHz}$=0.0066~cm$^2$~g$^{-1}$ as average
value between dust grains with thin and thick ice mantles for a volume density of $\sim$10$^5$ cm$^{-3}$
\citep{ossenkopf94} with an uncertainty of about a factor of 2. As a final step, we used the relationship
$A_\texttt{v}=1.258\times10^{-21}N_{\rm H_2}$ \citep{wagenblast89} to transform column density to visual extinction. The
resulting maps of the difference between the extinction maps derived from near-IR and mm data are shown in the right-hand
side panels of Fig.~\ref{fig:ext-emis}. For Core~40 we found an excess of extinction that could be due to the filtered
diffuse emission (see \S\ref{res_dust}). However, in such a case, one would expect this excess to be present over the
whole map. For Cores~14 and 48 we found a good agreement between both tracers. On the other hand, at  denser regions such
as the center of Core~109 ($n_{\rm H_2}$$>$4$\times$10$^5$~cm$^{-3}$), the $A_\texttt{v}$ derived from the 1.2~mm dust is
significantly larger, $\ga$10~mag, than that derived from the near-IR. This is the core with the highest column density
($\sim$4.8$\times$10$^{22}$~cm$^{-2}$), therefore this suggests that near-IR extinction maps constructed from 2MASS
catalogs do not have enough sensitivity or sampling scale to resolve the centers of very dense cores.  In such dense
regions, the number of 2MASS catalog background stars is not high enough to provide neither a large number of sources per
pixel, nor a large number of high extinction  measurements, thus the high extinction regions might be poorly resolved and
under-estimated.  These biases may explain, combined with the larger radius, the lower densities reported by
\cite{rathborne08}. Extensive observations toward the Perseus cloud in visual extinction and in radio continuum provide
similar results \citep{kirk06}. Extinction maps with higher resolution, made with deeper observations, are able to
resolve better the high extinction levels. For example,   \citet{kandori05} observed Core~109 (named FeSt~1-457) in
$A_{\texttt v}$ deeper with a resolution of $\sim$30$''$, and found a morphology in perfect agreement with our continuum
observations. Their $A_\texttt{v}$ intensity peak at the core center of $A_{\texttt v}$$\sim$41.0 (the largest in their
sample) is very close to our derivation, $A_{\texttt v}$$\sim$39.2, for a 30$''$ beam.

Summarizing, our dust continuum maps seem to be better at tracing the high extinction regions of the prestellar cores, at
least at this spacial resolution. These results suggest that the dust continuum  emission would trace the dense and cold
cores better than the 2MASS derived visual extinction. On the other hand, the visual extinction would be more  sensitive
to the cloud diffuse extended emission.


\subsection{Discussion on the individual cores}

\subsubsection{Core 14\label{dis_c14}}

Core~14, located in B59, is a compact and dense core but the less massive in our sample.  It is the only core that
belongs to a clumpy and filamentary structure, which is elongated along the NE-SW direction with an extent of
$\sim$500$''$ ($\sim$0.35~pc, see Fig.~\ref{fig:mambo}), with a morphology quite similar to that shown in $A_\texttt{v}$
maps \citep{rathborne08,roman09}. The location of Core~14 inside an elongated and clumpy filament, suggests that probably
it is still undergoing fragmentation, which could lead to the formation of smaller cores. In fact, it is resolved in
several small clumps which have sizes comparable to the sizes of the other cores in the Pipe, with radii of about
$\sim$0.04~pc.

Core~14 shows emission in all the early-time molecules at 3~mm.  CS~(2--1) and C$^{34}$S~(2--1) are clearly detected (see
Fig.~\ref{fig:lines1}), and the  abundances are the largest of the sample (see Table~\ref{tab_abun}). On the other hand,
C$_3$H$_2$~(2$_{1,2}$--1$_{1,0}$) and N$_2$H$^+$~(1--0) show weak emission and, consequently, low abundances.
\citet{rathborne08} detect weak NH$_3$ emission, in good agreement with  our N$_2$H$^+$ measurements. Only the main
component of the CN~(1--0) transitions is clearly detected. These are signatures of an object very young chemically.


\subsubsection{Core 40\label{dis_c40}}

Core~40, located in the {\it stem}, is another core with irregular morphology.  This  core shows emission  in all the
transitions at 3~mm, of both  early- and late-time molecules, and in the late-time N$_2$H$^+$ transition at 3~mm. The
molecular emission of Core~40 is strong, and only the emission of Core~109 is more intense, except for CS, which shows
the same $T_{\rm MB}$ for both cores.  These cores are the only ones that show strong late-time molecule emission.
Core~40 presents the highest CN and N$_2$H$^+$ abundances (see Table~\ref{tab_abun}). Regarding N$_2$H$^+$, the intense
emission with all the hyperfine components detected  is in perfect agreement with previous results of NH$_3$
\citep{rathborne08}. The HCN emission for Core~40 is quite anomalous, because the main hyperfine component is weaker than
the satellite components (see Fig.~\ref{fig:lineshfs}).  This suggests that the emission is not in LTE. 
\cite{gonzalez93} investigated with Monte-Carlo techniques the variation in HCN~(1--0) profiles.  According to their
work, an infalling cloud with a dense central core (see Fig.~\ref{fig:mambo}) surrounded by a  large diffuse envelope
\citep{lombardi06} may produce an HCN~(1--0) spectrum as the observed toward Core~40.


\subsubsection{Core 48\label{dis_c48}}

Core~48, located in the {\it stem}, has a quite elongated morphology. It is embedded in an environment with  high
polarization angle dispersion \citep{alves08}, which is the exception of this polarimetrically defined region.  It is
very diffuse, this is the largest and the less dense core in the sample. It shows emission only in three  early-time
molecules: CS~(2--1), C$^{34}$S~(2--1) and, marginally, HCN~(1--0).  The abundances of CS and C$^{34}$S are among the
largest in the sample, slightly lower than those for Core~14. The N$_2$H$^+$ molecule was undetected, in  agreement with
previous measurements of NH$_3$ \citep{rathborne08}.


\subsubsection{Core 109\label{dis_c109}}

Core~109, located in the {\it bowl}, is the most circular and compact core in the sample. The dust continuum emission of
this core is similar to that of the other cores. However, it is the densest one in our sample and the most massive
($\sim$4~$M_\odot$).  \citet{kandori05} find, through a Bonnor-Ebert profile fit, that Core~109 is gravitationally
unstable. \citet{aguti07} find, through observations of molecular transitions, that this core (designated also as
FeSt~1-457) is gravitationally bound.  \citet{kandori05} suggest other models, apart from a Bonnor-Ebert sphere,
including extra supporting mechanisms that might fit the density profile.  \citet{aguti07} propose that Core~109 is
pulsating, based on expansion motions of the outer layers. However, their Jeans mass measurement is compatible with the
mass of the core and they propose a quasi-stable state near hydrodynamic equilibrium. This core is embedded in a
magnetized medium (see \S~\ref{dis_cor}), thus, magnetic support could be a plausible source of external support.

This core shows emission of all the detected early- (C$_3$H$_2$, HCN, CS, C$^{34}$S, and CN) and late-time molecules
(N$_2$H$^+$, N$_2$D$^+$ and DCO$^+$).  The molecular emission of this core is always the strongest. Core~109 shows a very
strong N$_2$H$^+$ emission, in agreement with the NH$_3$  measurements by \citet{rathborne08}. As seen in
Table~\ref{tab_abun}, Core~109 has similar abundances for early-time molecules to those of the other cores.
Interestingly, the CS and C$^{34}$S abundances are  the lowest in our sample,  which suggests CS depletion toward the
center (detected on C$^{18}$O, \citealp{aguti07}).


\subsection{Qualitative chemistry analysis}
\label{dis_chem}


\begin{table*}[htbp]
\caption{
Molecular column densities of the chemical species observed toward the Pipe Nebula cores in cm$^{-2}$.
}
\begin{tabular}{c cccccccc}
\noalign{\smallskip}
\hline\hline\noalign{\smallskip}
\multicolumn{1}{c}{Source}
&\multicolumn{1}{c}{C$_3$H$_2$ $^a$}	&	\multicolumn{1}{c}{CS}
&\multicolumn{1}{c}{C$^{34}$S}		&	\multicolumn{1}{c}{CN}
&\multicolumn{1}{c}{HCN}		&	\multicolumn{1}{c}{N$_2$H$^+$}
&\multicolumn{1}{c}{N$_2$D$^+$ $^a$}	&	\multicolumn{1}{c}{DCO$^+$ $^a$}\\
\noalign{\smallskip}
\hline\noalign{\smallskip}
Core 14 &	$\phantom{< \ }3.84\times10^{11}$	& 	$\phantom{< \ }3.07\times10^{13}$ 
	&	$\phantom{< \ }6.19\times10^{11}$	&	$\phantom{< \ }1.16\times10^{12}$ 
	&	$<5.42\times10^{10}$ 		& 	$\phantom{< \ }9.70\times10^{10}$
	&	$<8.09\times10^{09}$	&	$<5.13\times10^{11}$	\\

Core 40 &	$\phantom{< \ }1.65\times10^{12}$ 	& 	$\phantom{< \ }7.11\times10^{12}$ 
	&	$\phantom{< \ }2.94\times10^{11}$	& 	$\phantom{< \ }4.70\times10^{12}$ 
	& 	$\phantom{< \ }2.57\times10^{12}$ 	& 	$\phantom{< \ }4.89\times10^{11}$ 
	&	$\phantom{< \ }2.17\times10^{09}$	&	$<4.93\times10^{10}$	\\

Core 48 & 	$<7.93\times10^{10}$ 	& 	$\phantom{< \ }1.53\times10^{13}$ 
	&	$\phantom{< \ }2.94\times10^{11}$	& 	$<1.64\times10^{11}$ 
	& 	$\phantom{< \ }2.59\times10^{12}$ 	& 	$<3.79\times10^{10}$ 
	&	$<6.88\times10^{09}$	&	$<7.82\times10^{10}$	\\

Core 109& 	$\phantom{< \ }6.36\times10^{12}$ 	& 	$\phantom{< \ }1.25\times10^{13}$ 
	&	$\phantom{< \ }3.68\times10^{11}$	& 	$\phantom{< \ }2.76\times10^{12}$ 
	& 	$\phantom{< \ }9.80\times10^{12}$ 	& 	$\phantom{< \ }6.79\times10^{11}$ 
	&	$\phantom{< \ }2.72\times10^{10}$	&	$\phantom{< \ }1.41\times10^{11}$	\\
\hline
\end{tabular}

(a) Transition with no opacity mesurements available, thus optically thin emission is assumed to obtain lower limits of
the column densities.
\label{tab_mol_col_dens}
\end{table*}


\begin{table*}[htbp]
\caption{
Abundances$^a$ of the chemical species with respect to H$_2$ observed toward the Pipe Nebula cores.
}
\begin{tabular}{c cccccccc}
\noalign{\smallskip}
\hline\hline\noalign{\smallskip}
\multicolumn{1}{c}{Source}& 
\multicolumn{1}{c}{C$_3$H$_2$ $^b$}	&	\multicolumn{1}{c}{CS}&
\multicolumn{1}{c}{C$^{34}$S}		&	\multicolumn{1}{c}{CN}& 
\multicolumn{1}{c}{HCN}			&	\multicolumn{1}{c}{N$_2$H$^+$}&
\multicolumn{1}{c}{N$_2$D$^+$ $^a$}	&	\multicolumn{1}{c}{DCO$^+$ $^a$}\\
\noalign{\smallskip}
\hline\noalign{\smallskip}
Core 14 & 	$\phantom{< \ }3.45\times10^{-11}$	& 	$2.77\times10^{-09}$ 
	&	$5.58\times10^{-11}$			&	$\phantom{< \ }9.58\times10^{-11}$ 
	& 	$<4.88\times10^{-12}$ 		& 	$\phantom{< \ }8.73\times10^{-12}$ 
	&	$<5.84\times10^{-13}$	&	$<2.93\times10^{-11}$	\\

Core 40 & 	$\phantom{< \ }1.55\times10^{-10}$	& 	$6.64\times10^{-10}$ 
	&	$2.75\times10^{-11}$			& 	$\phantom{< \ }4.19\times10^{-10}$ 
	& 	$\phantom{< \ }2.41\times10^{-10}$	& 	$\phantom{< \ }4.58\times10^{-11}$ 
	&	$\phantom{< \ }1.69\times10^{-13}$	&	$<3.73\times10^{-12}$	\\

Core 48 &	$<1.13\times10^{-11}$ 	& 	$2.19\times10^{-09}$
	&	$4.21\times10^{-11}$		& 	$<2.22\times10^{-11}$ 
	& 	$\phantom{< \ }3.71\times10^{-10}$& 	$<5.43\times10^{-12}$ 
	&	$<7.75\times10^{-13}$	&	$<7.17\times10^{-12}$		\\

Core 109& 	$\phantom{< \ }2.06\times10^{-10}$& 	$4.06\times10^{-10}$ 
	&	$1.19\times10^{-11}$	& 	$\phantom{< \ }8.54\times10^{-11}$ 
	& 	$\phantom{< \ }3.18\times10^{-10}$& 	$\phantom{< \ }2.20\times10^{-11}$ 
	&	$\phantom{< \ }7.30\times10^{-13}$&	$\phantom{< \ }3.37\times10^{-12}$	\\
\hline
\end{tabular}

(a) See Tables~\ref{tab_dust_col_dens} and \ref{tab_mol_col_dens} for dust and line column densities.\\
(b) Transition with no opacity mesurements available, thus optically thin emission is assumed to estimate a lower limit of
the column densities and, consequently, of the abundances.\\
(c) Due to the lack of C$^{34}$S data we assume optically thin emission to obtain a lower limit of the column density and,
as a result, also for the abundance.
\label{tab_abun}
\end{table*}



\begin{table*}[]
\caption{
Pipe Nebula core general properties with respect to core 109.
}
\begin{tabular}{crrrrrrrrrr}
\hline\hline
& \multicolumn{1}{c}{Diameter} 
& \multicolumn{1}{c}{Mass} 
& \multicolumn{1}{c}{$N_{\mathrm{H_2}}$} 
& \multicolumn{1}{c}{$n_{\mathrm{H_2}}$} 
& \multicolumn{1}{c}{$p_{\%}~^a$} 
& \multicolumn{1}{c}{$\delta PA~^a$} 
& \multicolumn{1}{c}{X(N$_2$H$^+$)} 
& \multicolumn{1}{c}{X(CN)} 
& \multicolumn{1}{c}{X(C$_3$H$_2$)} 
& \multicolumn{1}{c}{X(CS)} 
\\
& \multicolumn{1}{c}{(pc)} 
& \multicolumn{1}{c}{($M_{\odot}$)} 
& \multicolumn{1}{c}{(10$^{21}$cm$^{-2}$)} 
& \multicolumn{1}{c}{(10$^{4}$cm$^{-3}$)} 
& \multicolumn{1}{c}{(\%)} 
& \multicolumn{1}{c}{($^o$)} 
& \multicolumn{1}{c}{(10$^{-11}$)} 
& \multicolumn{1}{c}{(10$^{-11}$)} 
& \multicolumn{1}{c}{(10$^{-11}$)} 
& \multicolumn{1}{c}{(10$^{-11}$)} 
\\
\hline
Core 109& 0.063&  4.00& 47.60&  36.57&  11.0&  3.9& 2.20 & 8.54 &  20.6 & 40.6 \\
\\
\hline
  Relative values
& \multicolumn{1}{c}{Diameter} 
& \multicolumn{1}{c}{Mass} 
& \multicolumn{1}{c}{$N_{\mathrm{H_2}}$} 
& \multicolumn{1}{c}{$n_{\mathrm{H_2}}$} 
& \multicolumn{1}{c}{$p_{\%}$} 
& \multicolumn{1}{c}{$\delta$PA} 
& \multicolumn{1}{c}{X(N$_2$H$^+$)} 
& \multicolumn{1}{c}{X(CN)} 
& \multicolumn{1}{c}{X(C$_3$H$_2$)} 
& \multicolumn{1}{c}{X(CS)} 
\\
\hline
Core 109&      10.0 &     10.0 &     10.0 &     10.0 &  10.0& 10.0&       10.0   &        10.0  &       10.0   &      10.0 \\
Core 40 & 16.5& 6.3& 2.3& 1.4&   4.2& 22.2&  20.8  &   49.1 &   7.5  & 16.4 \\
Core 14 & 11.3& 3.5& 2.8& 2.5&   1.8& 40.4&   4.0  &   11.2 &        1.7   & 68.2 \\
Core 48 & 20.2& 5.2& 1.3& 0.6&   1.8& 83.8& $<$2.5 & $<$2.6 & $<$0.6 & 53.9 \\
\hline
\end{tabular}

(a) \citet{franco10}
\label{tab_resum}
\end{table*}


Table~\ref{tab_dust_col_dens} shows a variation  of about a factor of $\sim$4 around $10^{22}$\cmd\ of the average H$_2$
column densities derived for each of the cores with a 27$''$ beam, the one used to calculate the abundances for the
molecular transitions at 3~mm.  This represents, using  the relationship $A_V=6.289\times10^{-22}N_H$
\citep{wagenblast89}, average  values of A$_V$$\sim$4.4 to $\sim$19.4. The first case would represent a shallow core,
more affected by the external  radiation field, which tends to have a younger chemistry. The other extreme probably
indicates a denser and more shielded core, where one would  expect to find more complex and evolved molecules. However,
note that this also depends on the time-scale needed to form the core \citep{tafalla04,crapsi05}.

We find that CS (see Table~\ref{tab_abun}), an early-time molecule,  is detected in all the cores with abundances with
respect to H$_2$  of a few times $10^{-10}$, similar  to the ones found in other dense cores \citep{irvine87} or the ones
obtained  in gas--phase chemical models \citep{taylor96,garrod04}. It is worth to mention that Cores~14 and 48 show high
CS abundance, one order of magnitude higher than Cores~109 and 40. A similar result is found for the C$^{34}$S
abundances. The derived abundances for the early-time molecule HCN toward the cores in our sample is very uniform, and
seems to be independent of their physical properties.  The early-time molecule CN, a molecule that is also detected
commonly in dense cores, has also a significantly lower abundance (a factor $\ga$4) toward Core~48 than toward the rest
of the sample. Where detected, the CN abundance varies only within a factor of five. On the other hand, another
early-time molecule such as C$_3$H$_2$, shows differences in abundances of at least a factor of five among Cores~14 and
48 with respect to Cores~40 and 109. Late-time molecules, such as \ndh\ or deuterated molecules, are not broadly detected
in our sample: \ndh\ is detected except in Core~48. On the contrary, \ndd\ is only detected on Cores~40 and 109, and 
\dcop\ only in Core~109. We found a higher abundance of \ndh\ toward Core~40 than toward Core~109 by a factor of $\sim$2,
while \citet{rathborne08} found  an abundance of NH$_3$ toward Core~109 higher than that of Core~40 by a factor of
$\sim$3.4. However, both Cores~40 and 109 show higher abundances in \ndh\ than Cores~14 and 48.  Despite Cores~14 and 40
having a similar average column density, the former shows 5 times less abundance of \ndh\ than the latter. Moreover,
Core~14 does not show emission in any other late-time molecule while Core~40 is detected in \ndd\ showing an abundance
only a factor of 4 lower than that of Core~109. Briefly, the higher abundances in Cores~109 and 40 with respect to 
Core~14 and in particular to Core~48 (except for CS) is an indication that Cores~109 and 40 are  more chemically evolved
than Cores~14 and 48. However, the molecular abundances of these two late-time species are roughly an order of magnitude
lower than the prototypical starless cores L1517B and L1498 \citep{tafalla06}, which suggests that Cores~109 and 40 may
be in an earlier evolutionary stage than cores in Taurus.

\cite{rathborne08} observed the emission of the NH$_3$~(1,1),  NH$_3$~(2,2), CCS~(2$_1$--1$_0$), and HC$_5$N~(9--8)
transitions towards 46 cores of the Pipe Nebula. Cores 14, 40, 48, and 109 were included in their observations.  None of
the lines were detected in Core~48, which is shown to be again the more chemically poor core of our sample. HC$_5$N was
not detected in core 14, which also has the weakest CCS and NH$_3$ lines. The four transitions were detected in cores 40
and 109, but with some differences. The NH$_3$ lines are much more intense in core 109, a factor of $\sim$4 for the (1,1)
transition and $\sim$9 for the (2,2) line, while the CCS line is more intense in core 40, less than a factor of $\sim$2,
and the HC$_5$N lines are very similar in both cores, inside the RMS. All these results are consistent with our
observations: Core~48, which did not show emission of late-time molecules, is very poor chemically and shows a very young
chemistry. Core~14, has some very weak emission of late-time molecules (NH$_3$) but only weak emission of CCS, an
early-time molecule. Core 40 is more evolved chemically and shows stronger emission of early-time molecules than of
late-time molecules. Finally, Core~109 is the one showing more diversity of molecules and the more intense emission, in
particular of late-time molecules. Interestingly, the CCS abundance in core 109 is probably lower than in core 40, which
is consistent with the view that the CCS molecule is destroyed soon after the formation of a dense core, probably as a
result of the contraction of the core \citep{degregorio06,millar90,suzuki92}. This would reinforce the view that this
core is in a very advanced evolutionary state.

In summary, Core 109 seems to be the more chemically evolved core,  probably because it is more dense and because it
shows higher abundances of  late-time molecules. Core 40, with three times lower column density, also shows  large \ndh\
abundances. It might be in  an intermediate chemical evolutionary stage. These two cores probably are in an evolutionary
stage slightly younger than that of the prototypical starless  cores \citep{tafalla04, crapsi05}. Cores~48 and 14 show
similar physical properties in terms of size, mass and H$_2$ column density, to Cores~109 and 40. However, they appear to
be very chemically  poor and, therefore, they could be in an even younger stage of chemical evolution.


\subsection{Evolutionary trend and correlation with the diffuse gas}
\label{dis_cor}

Table~\ref{tab_resum} shows the summary of the main properties of the cores relative to Core 109, which is the one that
shows the strongest line emission. In this table we show the physical and chemical properties. Additionally, we added the
averaged polarimetric properties of the diffuse envelope around the cores \citep{alves08, franco10}: polarization
fraction ($p_\%$) and dispersion of the polarization position angle ($\delta PA$).  

As shown in Fig.~\ref{fig:mambo}, the polarization vectors calculated from optical extinction cannot be derived at the
more dense regions, where the visual extinction is higher. In Fig.~\ref{fig:mambo}, except for the map of Core~48 with
the lowest RMS, the polarization vectors lie in regions below the 3-$\sigma$ noise level. However, the trend of the
polarization vectors is in general  rather uniform over the whole map. Indeed, there are vectors up to very close to the
dense parts of the cores. Consequently, the derived magnetic field properties of the diffuse surrounding medium are also
representative of those of the dense part of the cores.

A relationship between the magnetic and the chemical properties of each core seems to exist.  The two more chemically
evolved cores, 109 and 40, appear to be embedded in a strongly magnetized environment, as $\delta PA$ values clearly
reflect (see Table~\ref{tab_resum}).  The other two cores, 14 and 48, do not show very different morphological properties
with respect to the previous two (size and mass). However, their chemical properties are completely opposed, and they are
likely younger cores in chemical time-scale. Interestingly, the magnetic properties of Cores~14 and 48 are also opposed
to those of Cores~40 and 109.  Cores~14 and 48 are surrounded by a molecular diffuse medium that is much more turbulent
than that surrounding the two previous ones. Core~14 is possibly affected by the star formation undergoing in the nearby
region B59. Core~48 appears to be dominated by turbulence and constitutes an exception in the \textit{stem}, whose cores
have uniform magnetic properties among them, showing low $p_\%$ and high $\delta PA$ \citep{franco10}. 

In summary, these four cores of the Pipe Nebula have similar masses and sizes, but they  are in different stages of
chemical evolution: Cores~109 and 40 are much more  evolved chemically than Cores~48 and 14. The different magnetic
properties of the  diffuse molecular environment suggest that Cores~109 and 40 have grown in a more quiescent and slowly
way (probably through ambipolar diffusion), whereas the growth of  Cores~14 and 48 has occurred much faster, an
indication that possibly a compression  wave that generates turbulence or the turbulence itself
\citep{falle02,ballesteros07}.  The longer time--scale of the ambipolar diffusion process could explain the more  evolved
chemistry found toward the cores surrounded by a magnetized medium. These features suggest two different formation
scenarios depending on the balance  between turbulent and magnetic energy in the surrounding environment. The importance 
of these results is worth of a more detailed study of the Pipe Nebula cores in order to  fully confirm these trends.


 \section{Summary and conclusions}

We carried out observations of continuum and line emission toward four starless cores of the Pipe Nebula spread out along
the whole cloud selected in base of their magnetic properties \citep{alves08, franco10}. We studied their physical
and chemical properties, and the correlation with the magnetic field properties of the surrounding diffuse gas.

\begin{enumerate}

\item The dust continuum emission of the observed Pipe Nebula cores shows quite different morphologies. In the sample there
are  diffuse cores, such as Cores~40 and 48, and compact and dense cores, such as Core~109. We have also mapped a clumpy
filament, which contains the embedded Core~14.  This filament is possibly undergoing  fragmentation into smaller cores of
sizes comparable to that of the others. We derived average radii  of $\sim$0.09~pc  ($\sim$18600~AU), densities of
$\sim$1.3$\times10^{5}$cm$^{-3}$, and  core masses of  $\sim$2.5~$M_\odot$.

\item The dust continuum peak coincides within the errors with $A_\texttt{v}$ peak derived from the  2MASS catalog.  The
continuum emission is more sensitive toward the dense regions, up to $\ga$10 magnitudes for the densest cores. On the
other hand, the diffuse emission is better traced by the extinction maps.  The masses are in average $\sim$3.4 times
smaller. 

\item We have observed several early- and late-time lines of molecular emission toward the cores and derived their column
densities and abundances.  The starless cores of the Pipe Nebula are all very young, but they present different chemical
properties possibly related to a different evolutionary stage. However, there does not seem to be a clear correlation
between the chemical evolutionary stage of the cores and their position in the cloud. Cores~109 and 40 show late-time
molecular emission and seem to be more chemically evolved.  Core~109 shows high abundances of late-time molecules and it
seems to be the more chemically evolved. Core 40 has three times lower H$_2$ column density than that of Core~109. It
presents a large \ndh\ abundance and the largest CN abundance, thus it might be in an intermediate chemical evolutionary
stage.  Cores~48 and 14 show only early-time molecular emission, and Core~14 presents weak \ndh\ emission, and seem to be
chemically younger than the other two cores. Core~14 has a similar mass and size than Core~40, but the N$_2$H$^+$,
C$_3$H$_2$, and CS abundances  are about one order of magnitude lower than the Core~40 abundances. Our results and
interpretation of the evolutionary stage of each core are consistent with the previous observations of \cite{rathborne08}
in these same cores.

\item There seems to be a relationship between the properties of the magnetic field in the cloud medium of the cores and the
chemical evolutionary stage of the cores themselves. The two more chemically evolved cores, 109 and 40, appear to be
embedded in a strongly magnetized environment, with a  turbulent to magnetic energy ratio of 0.05 and 0.27, respectively.
The two chemically younger cores, 14 and 48, appear to be embedded in a more turbulent medium. This suggests that the
magnetized cores probably grow in a more quiescent way, probably through ambipolar diffusion, in a time-scale large
enough to develop the richer chemistry found.  On the other hand, the less magnetized cores likely grow much faster,
probably in a turbulence dominated process,  in a time-scale too short to develop late-time chemistry.

\item The Pipe Nebula has revealed to be an excellent laboratory for the study of the very early stages of the star formation.
The studied cores show different morphologies, chemical evolutionary stages and magnetic properties.  The physical and
chemical properties are not directly linked as the competition between the magnetic field and  turbulence at small scales
seems to have an important influence in the core evolution. The importance of these results require a more detailed study
of the chemistry and magnetic field properties of the cores to  fully confirm these results.
\end{enumerate}

\begin{acknowledgements}

PF is supported by MICINN fellowship FPU (Spain). PF, JMG, MTB, JMM, FOA, GB, ASM, and RE are supported by MICINN grant
AYA2008-06189-C03 (Spain). PF, JMG, MTB, OM, FOA, and RE are also supported by AGAUR grant 2009SGR1172 (Catalonia). 
GAPF is partially supported by CNPQ (Brazil).
The authors want to
acknowledge all the IRAM 30-m staff for their hospitality during the observing runs, the operators and AoDs for their
active support, Guillermo Quintana-Lacaci for his help during the observing and reduction process of the bolometer
data, and Jens Kauffmann for helping on the implementation of his MAMBO-II new reduction scheme.

\end{acknowledgements}


\clearpage
\appendix
\section{A. Calculation of column density and mass of dust emission}
\label{app_dust}


\subsection{Radiative transfer equation and Planck function\label{sec_rad_trans}}

The intensity emitted by an assumed homogeneous medium of temperature
$T_{\mathrm{ex}}$ and optical depth $\tau_\nu$ at frequency $\nu$ is given by

\begin{equation}
I_\nu=
B_\nu(T_{\mathrm{ex}})\left(1-e^{-\tau_\nu}\right)
,
\label{eq_rad_trans}
\end{equation}

\noindent where $B_\nu$ is the Planck function,

\begin{equation}
B_\nu=
\frac{2h\nu^3}{c^2}\frac{1}{e^{h\nu/kT}-1}
,
\label{eq_planck}
\end{equation}

\noindent and $c$ is the speed of light, $k$ the Boltzmann's constant and $h$ the Planck's constant.
The Rayleigh-Jeans limit, $h\nu\ll kT$ (in practical units 
$
\left[\nu/\textrm{GHz}\right]
\ll
20.8 \left[T/\textrm{K}\right]
\label{eq_rj_limit}
$),
does not hold for MAMBO-II observations (250GHz) of prestellar cores ($T\simeq10$K), preventing the use of this
limit simplification.


\subsection{Telescope measurements\label{sec_tel}}

The beam solid angle is
$
\Omega_A=\int_{\mathrm{beam}}{Pd\Omega} 
\label{eq_beam}
$
\noindent
, where $P$ is the normalized power pattern of the telescope. Assuming that the telescope has a
gaussian beam profile, $P$ reads 
$
P(\theta)=
\exp(-4\ln2\theta^2/\theta^2_\mathrm{HPBW})
\label{eq_beam_profile}
$,
 where $\theta$ is the angular distance from the beam center. The beam solid angle is

\begin{equation}
\Omega_A=
\frac{\pi}{4\,\textrm{ln}(2)} \theta^2_{\mathrm{HPBW}}
.
\label{eq_beam_sangle}
\end{equation}

For discrete sources we measure flux densities, $S_\nu$, instead of intensities, $I_\nu$.
These two quantities are related by

\begin{equation}
S_\nu=
\int_{\mathrm{source}}{I_\nu \, P \, d\Omega}
.
\label{eq_flux_inten}
\end{equation}

\noindent This integration for a beam area, $S_\nu^{\rm beam}$, allows us to calculate
 the beam-averaged intensity as

\begin{equation}
\langle I_\nu\rangle =
\frac{S^{\rm beam}_\nu}{\Omega_A}
.
\label{eq_ave_int_beam}
\end{equation}


\subsection{From Flux to Column Density and Mass\label{sec_col_dens}}

One can calculate the opacity of the emission measured inside a beam, $\tau_\nu^{\rm beam}$,
 from equations~\ref{eq_rad_trans} and \ref{eq_ave_int_beam}, relating it with
the measured flux by

\begin{equation}
\tau^{\rm beam}_\nu=
-\textrm{ln}\left(
1-\frac{S^{\rm beam}_\nu}{\Omega_A \, B_\nu(T)}
\right)
.
\label{eq_tau_obs}
\end{equation}

On the other hand, the optical depth is defined as

\begin{equation}
\tau_\nu
\equiv
\int_\mathrm{line\ of\ sight}{\kappa_\nu \, \rho \, ds}
,
\label{eq_tau_def}
\end{equation}

\noindent where $\kappa_\nu$ is the absorption coefficient per unit density.

One can relate the column density with the optical depth and, thus, with the measured flux using

\begin{equation}
N_{\rm H_2}=
\int{n_{\rm H_2} \, ds}=
\int{\frac{\rho}{\mu \, m_\mathrm{H}}ds}=
\frac{\tau_\nu}{\mu \, m_\mathrm{H} \, \kappa_\nu}
,
\end{equation}

\noindent which particularized to a beam is
$
N^{\rm beam}_{\rm H_2}=\frac{\tau^{\rm beam}_\nu}{\mu \, m_\mathrm{H} \, \kappa_\nu}
,
\label{eq_col_dens_tau}
$,
where $m_\mathrm{H}$ is the hydrogen
mass and $\mu$ is the mean molecular mass per hydrogen atom.
In the case of optically thin emission
the intensity is proportional to the column density as
Eq.~\ref{eq_rad_trans} can be simplified to
$
I_\nu\approx B_\nu(T)\tau_\nu
\label{eq_rad_trans_simple}
$.

Then, the mass can be calculated as

\begin{equation}
M=
\mu \, m_\mathrm{H}\, D^2 \int{N_{H_2} \,d\Omega}
,
\end{equation}

\noindent which for a beam is
$
M^{\rm beam}=
\mu \, m_\mathrm{H}\, D^2 \, N^{\rm beam}_{H_2} \, \Omega_A
$,
where $D$ is the distance to the source. All these calculations can be 
applied to any solid angle bigger than a beam.

\section{B. Calculation of column density of line emission}
\label{app_line}


The column density for a $J \rightarrow J-1$ transition of a molecule ('Mol') is

\begin{eqnarray}
N_{\rm Mol}&=&
\frac{3\,k}{8\pi^3}
\frac{Q_{\rm rot}}{g_K\,g_I}
\frac{e^{Eu/T_{\rm ex}}}{\nu}
\frac{1}{S_{JkI}\,\mu^2}
\frac{J_{\nu}\left(T_{\rm ex}\right)}{J_{\nu}\left(T_{\rm ex}\right)-J_{\nu}\left(T_{\rm bg}\right)}
\frac{\tau}{1-e^{-\tau}}
\int_\mathrm{line}{T_{\rm MB} \, d\texttt{v}}
,
\end{eqnarray}

\noindent which translates into useful units as

\begin{eqnarray}
\left[\frac{N_{\rm Mol}}{\textrm{cm}^{-2}}\right]
&=&
1.67\times10^{14}\,
\frac{Q_{\rm rot}}{g_K\,g_I}
\left[\frac{S_{JkI}}{\textrm{erg\,cm}^3\,\textrm{statC}^{-2}\,\textrm{cm}^{-2}}\right]^{-1}
\left[\frac{\mu}{\textrm{D}}\right]^{-2}
e^{Eu/T_{\rm ex}}
\left[\frac{\nu}{\textrm{GHz}}\right]^{-1}
\frac{J_{\nu}\left(T_{\rm ex}\right)}{J_{\nu}\left(T_{\rm ex}\right)-J_{\nu}\left(T_{\rm bg}\right)}
\frac{\tau}{1-e^{-\tau}}
\left[\frac{\int_\mathrm{line}{T_{\rm MB} \, d\texttt{v}}}{\textrm{K\,km\,s}^{-1}}\right].
\nonumber
\\
\label{eq_col_dens}
\end{eqnarray}

\noindent Here, \noindent $J_{\nu}$ is the energy in units of temperature, and it reads
$
J_{\nu}\left(T\right)=
\frac{h\nu/k}{e^{h\nu/kT}-1}
$. See next subsections for a detailed description of all the terms involved.


\subsection{Observational terms}

\subsubsection{Single transitions\label{ssec_single}}

In case of single transitions, we have performed a Gaussian fit to the spectrum or a
statistical moment calculation, both using tasks from the CLASS package. We obtain 
from either analysis the main beam temperature, $T_{\rm MB}$,  the line velocity, $\texttt{v}$, and the 
integrated emission, $\int{T_{\rm MB} \, d\texttt{v}}$.

The opacity, $\tau$, is calculated numerically in those molecules with more than one transition observed.
 In the other cases we have assumed $\tau\sim0.3$.
The excitation temperature, $T_{\rm ex}$, can be calculated from the radiative transfer equation as

\begin{equation}
T_{\rm ex}=
\frac{h\nu}{k}
\left[\textrm{ln}\left(
\frac{h\nu/k}{\frac{T_{\rm MB}}{1-e^{-\tau}}+J_{\nu}\left(T_{\rm bg}\right)}+1
\right)\right]^{-1}
,
\label{eq_tex}
\end{equation}

\noindent where $T_{\mathrm{bg}}$ is the background temperature.


\subsubsection{Hyperfine transitions\label{ssec_h_fine}}

In case of hyperfine transitions, we take into account all the
hyperfine components of the selected transition. We have performed a hyperfine fit using CLASS, which
provides
$A\times\tau_m \, , \, \texttt{v}^{\rm reference \, line}_{\rm LSR} \, , \, \Delta \texttt{v} \, , \,
\tau_m$\, , where $A$ is

\begin{equation}
A=f\left(J_{\nu}(T_{\rm ex})-J_{\nu}(T_{\rm bg})\right)
,
\end{equation}

\noindent beeing $f$ is the filling factor assumed to be $\sim$1. \\

To be able to use equation \ref{eq_col_dens} as in the single transition case,
 we need $T_{\rm ex}$, $\tau$ and
$\int{T_{\rm MB} \, d\texttt{v}}$.
We can calculate $T_{\rm ex}$ as 
in equation \ref{eq_tex} calculating $T_{\rm MB}$ 
as $A\times\tau_m/\tau_m$, and $\tau_m$ is given by CLASS.
For the integrated emission, we can use

\begin{eqnarray}
\tau_0 \, \Delta \texttt{v}=
\int_\mathrm{line}{\tau \, d\texttt{v}}
\simeq\frac{1}{J_{\nu}(T_{\rm ex})-J_{\nu}(T_{\rm bg})}
\frac{\tau_0}{1-e^{-\tau_0}}
\int_\mathrm{line}{T_{\rm MB}(\texttt{v}) \, d\texttt{v}}
,
\end{eqnarray}

\noindent leading to

\begin{equation}
\int_\mathrm{line}{T_{\rm MB}(\texttt{v}) \, d\texttt{v}}\simeq
\tau_0 \, \Delta \texttt{v}
\left(J_{\nu}(T_{\rm ex})-J_{\nu}(T_{\rm bg})\right)
\frac{1-e^{-\tau_0}}{\tau_0}
.
\end{equation}

\noindent Making this transformations eq. \ref{eq_col_dens} can be used for hyperfine transitions.



\subsection{Non-observational terms}



\subsubsection{Partition function ($Q_{\rm rot}$)\label{sec_qrot}}

The rotational partition function, $Q_{\rm rot}(T)$, is defined as

\begin{equation}
Q_{\rm rot}(T)
\equiv
\sum g_{J} \, g_{k} \, g_{I} \, e^{-hBJ(J+1)/kT}
,
\label{eq_qrot}
\end{equation}

\noindent where the $g_X$ factors are the degeneration of the respective quantic number, in particular
$g_{J}=2J+1$.

Eq.~\ref{eq_qrot} can be approximated, in the limit of high temperatures, by an integral because generally
the energy levels are close together.
We are only interested in the high temperature limit because is when the
transition is activated, so this limit is accurate enough.\\

\noindent $\bullet$ {\bf Linear molecules:}
The solution for the diatomic case is general for any lineal molecule, so long as the molecular moment of
inertia is computed properly for more than 2 atoms.

For lineal molecules $g_{k}=1$, $g_{I}=1$ and $g_J=(2J+1)/\sigma$. $\sigma$ (the symmetry number) is 1 for
heteronuclear diatomic (C-O) or asymmetric linear polyatomic (O-N-N) molecules, and 2 for homonuclear
diatomic (H-H) or symmetric linear polyatomic (O-C-O) molecules.

The partition function at high temperatures can be calculated as
\begin{eqnarray}
Q_{\rm rot}
 \simeq 
\frac{1}{\sigma}\int^\infty_0{(2J+1)e^{-hBJ(J+1)/kT}dJ}
 \simeq 
\frac{1}{\sigma}\int^\infty_0{e^{-(J^2+J)hB/kT}d(J^2+J))}
\simeq \frac{1}{\sigma}\frac{k \, T}{h \, B}
,
\end{eqnarray}

\noindent where $B$ is the rotational constant available at the catalogues.
A more accurate expression (Pickett et al., 2002) used in this work is
\begin{eqnarray}
Q_{\rm rot}\approx\frac{1}{\sigma}\left(\frac{k \, T}{h \, B}+\frac{1}{3}+\frac{1}{15}\frac{\sigma  \, h \, B}{k \, T}+...\right)
.
\end{eqnarray}

\noindent $\bullet$ {\bf Non-linear molecules:}
Non-linear molecules have up to three moments of inertia and, thus, three rotational
constants ($A, B, C$). In a similar way than before, but more complicated,
the calculation of the rotational partition function at high temperatures is

\begin{equation}
Q_{\rm rot}\approx\frac{\sqrt{\pi}}{\sigma}\left(\frac{K \, T}{h}\right)^{3/2}\frac{1}{\sqrt{A \, B \, C}}
.
\end{equation}


\subsubsection{Upper level energy ($E_u$)\label{ssec_eu}}

We can calculate the energy of the upper level ($E_u$) as a function of the lower
level ($E_l$) plus the energy of the photon emitted (both available at catalogues). This is, 
in units of temperature and using the units given in the catalogues,

\begin{equation}
\left[\frac{E_u}{\textrm{K}}\right]=
1.4388
\left[\frac{E_l}{\textrm{cm}^{-1}}\right]
+
4.799\times10^{-5}
\left[\frac{\nu}{\textrm{MHz}}\right]
.
\end{equation}


\subsubsection{Intrinsic line strength times squared dipolar momentum ($S\,\mu^2$)\label{ssec_smu2}}

We can calculate the product of the \emph{intrinsic} line strength, $S_{JkI}$, and
the squared dipolar momentum, $\mu^2$, from the $Q_{\rm rot}$ at 300~K ($Q^{300}_{\mathrm{rot}}$), 
the line strength (LogINT) at 300~K
and the lower state energy ($E_l$). All these parameters are available at the catalogues.

In a usable form,

\begin{eqnarray}
\left[\frac{S\mu^2}{\textrm{erg\,cm}^3\,\textrm{statC}^{-2}\,\textrm{cm}^{-2}\,\textrm{D}^{-2}}\right]
=
24025
\times\left[\frac{10^{LogINT}}{\textrm{MHz\,nm}^2}\right]\;Q^{300}_{\mathrm{rot}}
\left[\frac{\nu}{\textrm{MHz}}\right]^{-1}
\left(\mathrm{exp}
	\left\{
	4.796\times10^{-3}
	\left[\frac{E_l}{\textrm{cm}^{-1}}\right]
	\right\}
\right)
\left(1-\mathrm{exp}\left\{
	-1.6\times10^{-7}
	\left[\frac{\nu}{\textrm{MHz}}\right]
\right\}\right)^{-1}.
\nonumber
\\
\end{eqnarray}

	       
\end{document}